\documentclass[apj,twocolumn]{openjournal}

\usepackage[T1]{fontenc}
\usepackage[breaklinks,colorlinks,citecolor=blue,urlcolor=blue]{hyperref}
\usepackage{natbib}

\usepackage{xcolor}

\usepackage{xspace}
\usepackage{wrapfig}
\newcommand{\kNNKDE}{$k$NN$\times$KDE\xspace}
\newcommand{\MJ}{M$_{\rm J}$\xspace}
\newcommand{\RJ}{R$_{\rm J}$\xspace}
\newcommand{\ME}{M$_\oplus$\xspace}
\newcommand{\RE}{R$_{\oplus}$\xspace}

\newcommand{\orcidauthor}[3]{\author{\href{http://orcid.org/#1}{#2$^{#3}$}}}

\usepackage{soul} 

\begin{document}

\title{\vspace{-0.8cm} Estimating Exoplanet Mass using Machine Learning on Incomplete Datasets\vspace{-1.5cm}}

\orcidauthor{0000-0001-6676-1484}{Florian Lalande}{1,*}
\orcidauthor{0000-0001-6692-612X}{Elizabeth Tasker}{2}
\orcidauthor{0000-0002-2446-6820}{Kenji Doya}{1}

\affiliation{$^{1}$Okinawa Institute of Science and Technology. 1919-1 Tancha, Onna, Kunigami, Okinawa 904-0495, Japan}

\affiliation{$^{2}$Institute of Space and Astronautical Science, JAXA, Yoshinodai 3-1-1, Sagamihara, Kanagawa 252-5210, Japan}

\thanks{$^*$E-mail: \href{mailto:florian.lalande@oist.jp}{florian.lalande@oist.jp}}

\begin{abstract}
The exoplanet archive is an incredible resource of information on the properties of discovered extrasolar planets, but statistical analysis has been limited by the number of missing values. One of the most informative bulk properties is planet mass, which is particularly challenging to measure with more than 70\% of discovered planets with no measured value. We compare the capabilities of five different machine learning algorithms that can utilize multidimensional incomplete datasets to estimate missing properties for imputing planet mass. The results are compared when using a partial subset of the archive with a complete set of six planet properties, and where all planet discoveries are leveraged in an incomplete set of six and eight planet properties. We find that imputation results improve with more data even when the additional data is incomplete, and allows a mass prediction for any planet regardless of which properties are known. Our favored algorithm is the newly developed \kNNKDE, which can return a probability distribution for the imputed properties. The shape of this distribution can indicate the algorithm's level of confidence, and also inform on the underlying demographics of the exoplanet population. We demonstrate how the distributions can be interpreted with a series of examples for planets where the discovery was made with either the transit method, or radial velocity method. Finally, we test the generative capability of the \kNNKDE to create a large synthetic population of planets based on the archive, and identify potential categories of planets from groups of properties in the multidimensional space. All codes are Open Source\footnote{\href{https://github.com/DeltaFloflo/exoplanet\_imputation}{https://github.com/DeltaFloflo/exoplanet\_imputation}}.
\end{abstract}


\section{Introduction}
\label{sec:introduction}

Since the first discoveries in the early 1990s, over 5,500 planets have been discovered outside our Solar System \citep{Wolszcan1992, Mayor1995, Akeson2013}. While the planets orbiting our Sun can be categorized as either rocky or gaseous simply depending on their orbital period, the myriad of sizes and orbits of the planets detected around other stars point to a multitude of formation pathways for planets that are influenced by a wide range of environmental factors.

Dedicated survey missions such as Convection, Rotation et Transits planétaires (CoRoT), the Kepler space telescope, and the Transiting Exoplanet Survey Satellite (TESS), alongside ground-based search instruments and programs that include the High Accuracy Radial Velocity Planet Searcher (HARPS), Wide Angle Search for Planets (WASP) and Optical Gravitational Lensing Experiment (OGLE) are trying to build a census of planet types. This has resulted in the construction of a large archive of data for the properties of the discovered planets. This exoplanet archive is an invaluable resource for identifying patterns and trends between planet and stellar properties. Such relationships can be used to estimate properties of planets that have not (and often cannot) be measured, allowing a more complete picture of planet diversity and information to select the most promising targets for time-consuming atmospheric characterization studies by instruments such as the James Webb Space Telescope. However, making full use of the archive has turned out to be challenging.

One of the principal difficulties is that the archive consists of thousands of planets, but each entry has recorded values for only a small subset of the measurable properties that varies depending on the discovery technique. For example, planets detected via the transit technique will have a measured radius, while the radial velocity technique provides a measured minimum mass. Similarly, direct imaging and gravitational microlensing techniques, which are sensitive to planets further from the host star, can measure a planet mass but include less orbital information due to being a single (or small section) snapshot of the planet trajectory. Host star properties are likewise sparsely measured, but their size and composition are expected to strongly impact the planet formation process \citep[e.g.][]{Fischer2005, Dressing2013, Osborn2020, Adibekyan2021, Bryant2023}. Detecting the same planet through multiple techniques can help fill these gaps, but is often not possible. A transit observation requires the planet to pass across the star's surface as observed from Earth; an alignment that gets proportionally less likely for longer orbits (geometric probability decreasing as the inverse of the orbital radius, $p_T = R_{\star}/a$, for stellar radius $R_{\star}$ and average orbital distance, $a$). Likewise, stellar activity (such as star spots) is a continual bane for radial velocity detection, microlensing requires a one-off chance alignment with a background star, and imaging is currently most sensitive to very distant, young massive planets \citep{Oshagh2017, Wallace2019}. The result is a large but sparse data archive of discovered planets, which is difficult to leverage to identify trends that simultaneously depend on multiple properties. 

For this reason, attempts to construct relationships between planet properties are usually based on a small subsection of the discovered planets and involve just two properties, such as planet mass and radius, planet occurrence as a function of mass and stellar metallicity, or planet multiplicity and orbital eccentricity \citep{Johnson2010, Weiss2014, Rogers2015, Limbach2015, Wang2015, ChenKipping2017, Petigura2018, Otegi2020}. But inevitably, two-dimensional relationships cannot capture evolution pathways that depend on multiple factors, and nor can they utilize planets in the archive that lack either of the considered properties. The first issue can make it difficult to determine when a trend exists due to noise from other dependent parameters that have not been considered (see also section~\ref{sec:data} and Figure~\ref{fig:pairplot}). The latter point reduces the number of examples that can be included in the data analysis, and also risks restricting studies to planets with particular properties in common (such as planets on short orbits which transit to provide a radius measurements) and resulting conclusions may not hold more broadly through the exoplanet population.

The issue of multiple dependencies has recently been tackled by the use of machine learning methods \citep{UlmerMoll2019, Tasker2020, MousaviSadr2023}. A strength of machine learning techniques is that multidimensional dependencies can be easily leveraged in a dataset, allowing more complex trends to be utilized when predicting planet properties. \citet{UlmerMoll2019} used a Random Forest regression model (see also section~\ref{sec:method}) while \citet{MousaviSadr2023} tested a series of regression algorithms to estimate planet radius based on planet, stellar and orbital properties. Both groups found that multiple parameters were important in estimating planet radius, with \citet{UlmerMoll2019} citing planet mass and equilibrium temperature as key values, and \citet{MousaviSadr2023} listing planet mass, orbital period and one stellar parameter out of stellar mass, stellar radius, or the stellar effective temperature as the most informative combination. The precursor to this research was \citet{Tasker2020} (hereafter TLG2020), which developed a neural network to impute missing values in the exoplanet archive. TLG2020 focused primarily on imputing missing planet mass and radius values, as the resulting average density is the most informative bulk property when considering planet composition or surface conditions \citep[e.g.][]{Bond2010, Unterborn2018, Bonomo2019}. The accuracy of the TLG2020 neural network--a modified Boltzmann Machine (mBM)--was slightly better than two-dimensional imputations, and covered a wide range of masses. Additionally, the mBM produced a relative likelihood function for the planet mass or radius that could be sampled to produce a probability distribution. This was an informative way to explore the imputation, with peaks indicating when multiple planet sizes could be found at a particular orbit in similar planetary systems. 

However, these machine learning implementations had a frustrating limitation. The algorithms had to be trained on data where every entry had a complete set of properties, with no missing values. This significantly limited how much of the exoplanet archive could be used by the mBM to find the multidimensional connections between properties. In order to have a dataset large enough to be used for machine learning, TLG2020 restricted the included properties to planet mass, planet radius, orbital period, equilibrium temperature, stellar mass, and the number of known planets in the system. This resulted in a dataset of 550 planets with six properties. Other properties were sufficiently sparsely measured that their inclusion would have reduced the dataset size too significantly for meaningful results. \citet{UlmerMoll2019} and \citet{MousaviSadr2023} were similarly limited to small datasets of between 500 - 700 planets.

In this paper, the limitation of complete dataset training is tackled by testing the ability of five different machine learning methods, all of which can leverage incomplete, multi-property datasets to impute missing information. This allows each algorithm to work with the data from all currently known planets, rather than a small subset that have the required measured properties. The algorithms can estimate any missing property in the dataset, but we focus the analysis on the imputation of planet mass. Planet mass is one of the most informative bulk properties for assessing the planet environment, as it is needed to determine both gravity and average density. However, mass is also one of the most challenging planet properties to measure through observation, with a 72.8\% missing rate in the exoplanet archive. This makes it an important use case of any imputation technique.

The algorithms can estimate the missing properties for any planet, regardless of which properties have been measured from observation. This removes restrictions such as requiring the planet radius in order to impute a planet mass by leveraging the other known planet properties. The performance when imputing values for planets with a smaller or larger number of observed values will be discussed in the results in section~\ref{sec:impute}.

Our favored algorithm is the newly developed \kNNKDE, which can return a probability distribution of the imputed value rather than a point estimate. The imputed mass distribution can be combined with a minimum mass measurement returned from radial velocity observations to improve the accuracy of the result. The distribution shape also indicates the algorithm's level of confidence in the imputed value, and can reveal information about the demographics of the underlying planet population that is hard to decipher from two-dimensional trends. For example, multiple peaks in the mass distribution can indicate that several planet sizes are commonly found with the same properties that have been observed. Examples of the mass distributions are analysed in section~\ref{sec:impute} to understand the imputation process so that the algorithm is not treated as a black box, and also improve the final imputed estimate.

The paper is structured as follows: section~\ref{sec:data} examines the exoplanet archive data, and the visible two-dimensional trends between planet and stellar properties. These underscore the need to utilize multidimensional relationships in the imputation of missing values, but are also important when discussing the results to understand the origin of the imputation. The five algorithms for imputing missing values are described in section~\ref{sec:method}. The results in section~\ref{sec:impute} are divided into three sub-sections, each using a different dataset to impute missing values. Section~\ref{sec:completedata} uses the same complete dataset with six properties as TLG2020, and compares the results with the mBM neural network. The dataset is then extended in section~\ref{sec:fullarchive} to include all planets in the archive, regardless of whether all of the six properties have measured values. This assesses the benefit of using a large but incomplete dataset, versus a small dataset but with complete set of properties. The final dataset in section~\ref{sec:extended} adds two more planet properties into the imputation and compares the result of this extra information. Finally, the \kNNKDE algorithm is used in section~\ref{sec:generative} as a generative model to create a large population of simulated planets based on the final dataset. Groups in the eight-dimensional property space are analysed as potential categories of planet evolution. A discussion of these results is in section~\ref{sec:conclusions}.

\section{The exoplanet data archive}
\label{sec:data}

\begin{figure*}[!ht]
    \centering
    \includegraphics[width=0.95\textwidth]{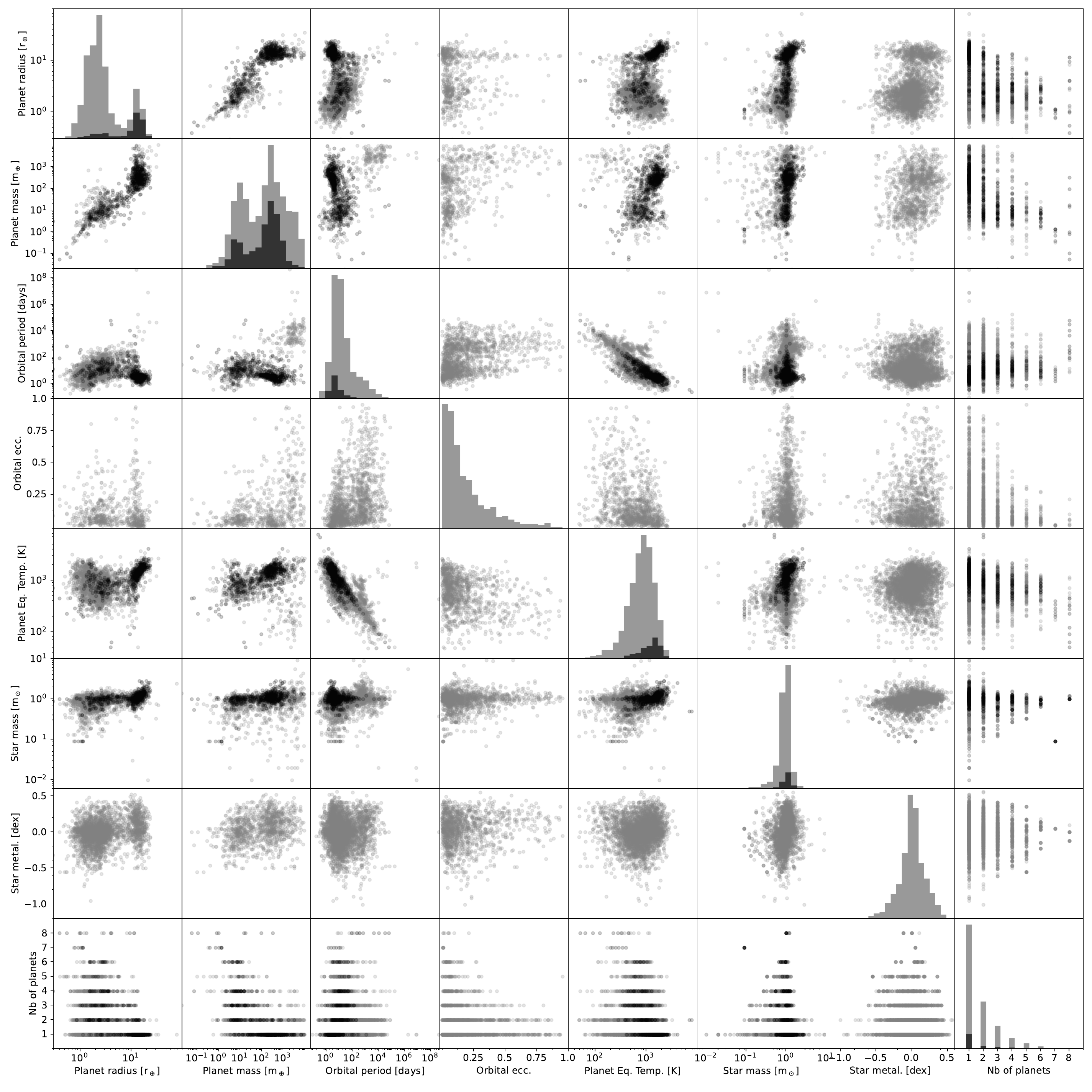}
    \caption{Pairplot for eight planet properties. Grey dots and histogram bars denote the full NASA Exoplanet Archive, while black dots represent the subset of planets used in the complete six properties dataset of TLG2020. Five variables (planet radius, planet mass, planet orbital period, planet equilibrium temperature, and stellar mass) have been log-transformed and are used as such in this study.}
    \label{fig:pairplot}
\end{figure*}

Each of the algorithms in this paper utilizes a dataset of planet properties to impute missing values. Of course, this is only successful if the values in the dataset are related such that the known properties for a planet can provide information on the unknown values. As mentioned in the introduction, identifying complex relationships between multiple properties is challenging, but a sense of whether pairs of planet properties are related can be gained by looking at the two-dimensional pairplots. The pairplots are shown in Figure~\ref{fig:pairplot} for eight planet properties. Diagonal panels show the univariate distributions as histograms, and off-diagonal panels show bivariate distributions as scatter plots. Throughout our work, we use the logarithm of five variables: planet radius, planet mass, planet orbital period, planet equilibrium temperature, and host star mass.

The dataset for exoplanet properties used in this work is the NASA Exoplanet Archive\footnote{\url{https://exoplanetarchive.ipac.caltech.edu/}}. Three different subsets of the full archive are explored in section~\ref{sec:impute} for imputing missing values. The first is the same dataset used in TLG2020 (downloaded from the NASA Exoplanet Archive in 2018. Dataset: \citet{TLG2020Data}), consisting of 550 planets each with six known properties: planet mass, planet radius, orbital period, planet equilibrium temperature, stellar mass and number of known planets in the system. The relatively small size of this dataset was due to the necessity in TLG2020 to use a dataset with complete properties and no missing values. The observed values for the planet properties in this complete dataset are shown in Figure~\ref{fig:pairplot} as black dots and histogram bars. 

The next two datasets utilized in section~\ref{sec:impute} use all 5,243 confirmed planets in the NASA Exoplanet Archive (as of February 2, 2023). The majority of planets in these datasets have incomplete properties, which can be handled by the algorithms used in this work. The first of these datasets uses the same six planet properties as in the complete complete of TLG2020, while the second dataset additionally adds orbital eccentricity and stellar metallicity. The observed values in these datasets are shown as grey dots and histogram bars in Figure~\ref{fig:pairplot}. All three datasets also include the eight planets of our Solar System (bringing the total number of planets in the second two datasets up to 5,251).

From the pairplot in Figure~\ref{fig:pairplot}, a number of trends between pairs of variables are apparent. Unsurprisingly, planet radius is positively correlated with the planet mass, while the orbital period of a planet is negatively correlated with the planet equilibrium temperature. Notably, even these relationships show significant scatter, indicating other factors are playing a role. For example, planets on short orbits may have inflated atmospheres that result in a higher planet radius for a given planet mass, and stellar type obviously influences the equilibrium temperature on a given orbital period.

Looking at the planet mass and radius, it can also be seen that the subset of planets that was used in TLG2020 (marked in black) when a dataset of complete properties with no missing values was required, is not representative of the full population of confirmed exoplanets. In particular, comparison of the gray and black histogram for planet radius in Figure~\ref{fig:pairplot} (top left) shows that the majority of super Earths with radius $1\,r_\oplus < r < 5\,r_\oplus$ are missing from the complete properties dataset, despite being a dominant population in the archive overall. The more easily observed transiting hot Jupiters are therefore over represented in this dataset, which risks a strong bias in the resulting imputation.

Hints at less strong dependencies can also be seen in Figure~\ref{fig:pairplot}. For example, planets in multi-planet systems tend to have smaller masses and radii. A similar trend was previously noted by \citet{Weiss2018}, who found that multi-planet systems discovered by Kepler often consisted of small ($r_p < 4$\,\RE), similarly sized planets (dubbed ``peas in a pod''). Massive gas giants on short orbits are also most likely to have migrated inwards, potentially disrupting other planets forming in the system by removing a substantial fraction of the planetesimal building material that would otherwise create rocky planets \citep{Armitage2003, Raymond2005, Raymond2006}. There is also likely an observational bias component here, as planets are more difficult to detect further from the star, so systems like our own with multiple cool gas giants, or those with smaller single planets orbiting further out, are not easily observable. Stellar mass shows a small trend with very large planet radii, although no notable pattern with smaller mass planets, which was also noted in regression models by \citet{MousaviSadr2023}. This may also be the effect of observational bias, since it is more challenging to find smaller planets around more massive stars, which will have a large ratio in their relative sizes.

The two additional parameters added to the third dataset in this study were orbital eccentricity and stellar metallicity. Evidence of trends between these properties and the original six planet properties can be seen, indicating that the addition has the potential to provide informative content. For example, eccentric orbits are more common for longer periods, as tidal interaction will act to reduce eccentricity at small separations from the host star. More interestingly, highly eccentric orbits are dominated by higher mass planets, and single planet systems. Planet multiplicity has previously been noted to correspond to low eccentricity, and it has been suggested that systems with a higher number of planets have suffered from less dynamical instability in the past, which would drive eccentric orbits \citep{Limbach2015, Ghosh2024}. Meanwhile, massive planets may drive dynamic instability after the evaporation of the protoplanetary disc. In systems with multiple gas giants, this can lead to planet scattering that places one planet on an eccentric orbit and the other ejected out of the system (or on a distance orbit that is more difficult to detect) \citep{Rasio1996, Juric2008}. This is possibly supported by a weak trend between stellar metallicity and orbital eccentricity, and between stellar metallicity and planet mass and radius. Metal rich stars are more likely to host massive planets, which in turn, may become dynamically unstable and create high eccentric orbits \citep{Dawson2013, Buchhave2018}. Eccentricity also usually requires a radial velocity detection, and this is an easier detection for high mass planets, introducing a bias towards finding more massive planets on less circular orbits.

\begin{figure}
    \centering
    \includegraphics[width=1.05\columnwidth]{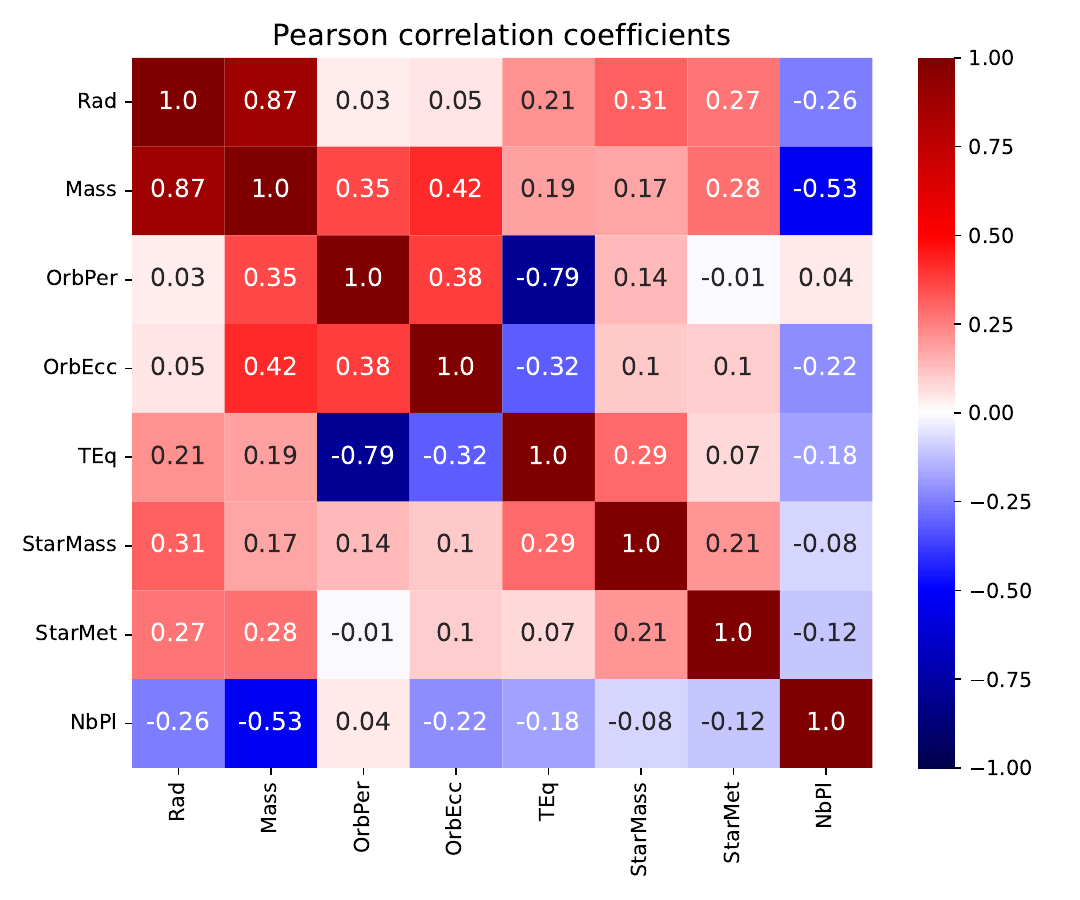}
    \caption{Pearson correlation coefficients for the eight chosen planet properties in the extended dataset. These values quantify the direction and the magnitude of pair-wise linear relationships between the exoplanet properties. Note that there exist no systematic way for interpreting Pearson correlation coefficients.}
    \label{fig:pearson_corr_coeff}
\end{figure}

An attempt to quantify the trends between pairs of variables can be made by using the Pearson correlation coefficients, shown in Figure~\ref{fig:pearson_corr_coeff}. The Pearson coefficient is a statistical measure of the linear correlation between two variables, taking a value between $\pm 1$ \citep{Pearson1895}. The magnitude of the coefficient indicates the strength of the correlation, with a $+1$ or $-1$ value corresponding to a perfectly linear relationship, as can be seen along the diagonal of Figure~\ref{fig:pearson_corr_coeff} between identical properties. A coefficient value of 0 indicates that there is no linear dependency between the two variables. There is no objective rule to interpret Pearson correlation coefficients, but a general rule of thumb suggests that magnitudes between $0.2 - 0.5$ indicate moderate linear correlations, and magnitudes above 0.5 indicate strong linear corrections.

The two largest correlation coefficients in Figure~\ref{fig:pearson_corr_coeff} unsurprisingly correspond to the strongest visible trends in the pairplot in Figure~\ref{fig:pairplot}. Planet radius and planet mass have a correlation coefficient of $0.87$, indicating the very strong positive correlation between these two measurements of planet size. And the next largest is the negative correlation coefficient $-0.79$, marking the inverse relation between planet equilibrium temperature and orbital period. Weaker trends visible in the pairplot are also seen here at smaller coefficient values. Planetary systems with a high number of known planets correlate fairly strongly with lower planet masses, with a coefficient of $-0.53$. Stellar mass and planet radius have a weaker correlation of $0.31$, reflecting that this trend is not apparent at all radii.

The correlation coefficients for orbital eccentricity and stellar metallicity support the evidence in the pairplot that the addition of these two properties should be important for predicting other values. A moderate coefficient value of $0.42$ can be found between orbital eccentricity and planet mass, and slightly weaker values of $0.38, -0.32$ and $-0.22$ between orbital period, equilibrium temperature, and number of planets in the system respectively. Stellar metallicity also has an indicated correlation with planet mass and radius (coefficient values of $0.28$ and $0.27$), supporting the trend that higher metallicity stars tend to host larger planets.

The multidimensional dependence of the planet properties is also clear from Figure~\ref{fig:pearson_corr_coeff}. While a few combinations have very low correlation values with magnitudes less than $0.1$, many pairs of properties show evidence of a moderate correlation. Similarly, very few pairs have extremely strong correlations, indicating that other processes are at play, and no single property is determining the value of another. This supports the idea that machine learning algorithms--with their ability to handle complex and highly dimensional dependencies--is a concept worth exploring in order to unlock the full potential of the exoplanet archive data. While an incomplete picture of the dependencies between properties, the trends noted here can also help understand the origin of the imputed values. These will be revisited in section~\ref{sec:impute}.

\subsection{Data cleaning}
\label{sec:clean}

While as many observed values as possible for the considered six or eight properties were used in constructing the datasets utilized by the algorithms, values that risked being misleading were removed. In particular, values labeled as limits (suggesting the true value might be significantly larger or smaller) or stated without valid error measurements (indicating the value might not be an observed measurement) were not included in the final datasets. Significance cuts (where high error values are removed from the dataset) was not included here, due to the risk of biasing the dataset towards larger planets where properties are easier to measure \citep{Burt2018}. (The latter is of course not completely removable, as a significance cut will occur due to the certainty needed for publication of a discovered property.)

It is worth noting that although this value is not presented as a lower limit, the number of planets in the system is inevitably the planets that have been discovered to date, not the true number of planets that may orbit that star. Despite this, the discovered trends discussed above do suggest that the number of known planets can be informative, such as the Kepler multi-planet systems often hosting smaller worlds, and the hot Jupiter disruption of planet formation \citep{Raymond2006, Weiss2018} (although the former may yet be a product of bias, as a higher mass system may have scattered planets on inclined orbits that do not transit). The number of known planets is therefore included in information given to the codes, which also maintains consistency with TLG2020 to allow the comparison with previous work.

Also as in TLG2020, missing planet equilibrium temperatures have been calculated using measurements of the host star radius $R_*$, the host star effective temperature $T_*$, and the average orbital distance $a$ when available via $T_{\rm eq.} = T_* \sqrt{R_*/(2a)}$. The derived value does have the potential to strongly overlap with the other considered properties, as the orbital distance will correlate with the period in the case of a circular orbit, while the stellar properties are frequently not measured independently from one another. However, since new information is available from this field in many cases and the host star radius $R_*$ is not used as it is, this property is still considered worth including. How overlap between properties is handled will depend on the algorithms. 

As certain planets have been observed on multiple occasions, the NASA Exoplanet Archive offers several sets of observed parameters. Generally, the archive default dataset is used in this study. However, for missing values where another study has proposed a trustworthy (not a limit or without error bars) measurement, this value is included.

\begin{deluxetable*}{llccc}[!ht]
  \tablecaption{The missing rate (percentage of planets without a measured value) for exoplanet properties considered in this study in the NASA Exoplanet Archive. Most notably, 72.8\,\% of the known planets do not have a measured mass.}
  \tablehead{
  \colhead{Property} & \colhead{Units} & \colhead{Minimum} & \colhead{Maximum} & \colhead{Missing rate}
  }
  \startdata
  Planet radius                    & $r_\oplus$  & $0.3$    & $77.3$         & $30.4\,\%$  \\
  Planet mass                      & $m_\oplus$  & $0.02$   & $9,852.0$      & $72.8\,\%$  \\
  Planet orbital period            & days        & $0.1$    & $402,000,000$  & $3.7\,\%$   \\
  Planet orbital eccentricity      & .           & $0.00$   & $0.95$         & $70.1\,\%$  \\
  Planet equilibrium temperature   & $K$         & $48$     & $7,719$        & $13.5\,\%$  \\
  Host star mass                   & $m_\odot$   & $0.01$   & $10.94$        & $0.5\,\%$   \\
  Host star metallicity            & dex         & $-1.00$  & $0.56$         & $10.1\,\%$  \\
  Number of planets in the system  & .           & $1$      & $8$            & $0.0\,\%$   \\
  \enddata
\label{table:missingrates}
\end{deluxetable*}

Table~\ref{table:missingrates} shows the chosen eight exoplanet properties considered in this study, and their missing rate, which is the fraction of planets that do not have a measured value. Since exoplanets have been discovered through various detection methods which provide different measured properties, the rate of missing values varies strongly across planet properties. As discussed in section~\ref{sec:introduction}, planet mass has a particularly high missing rate, since mass is not measured by the prolific transit technique (which accounts for roughly 75\,\% of planet detections), and only a minimum mass is measured by the radial velocity technique (approximately 20\,\% of detections). Measurements of the planet mass must therefore come either from a dual transit and radial velocity detection, or from the less common detection techniques, such as imaging or gravitational lensing. This paper therefore focuses on the accuracy of imputing planet mass, although the algorithms used in sections~\ref{sec:completedata} and \ref{sec:fullarchive} impute all properties that are missing.

\subsubsection{A note on bias}

Since the data included in this study is directly from the observed values, it will be impacted by both the intrinsic bias and sensitivity limits of the exoplanet detection techniques. These observational biases differ between each technique that has added planet discoveries to the archive, and no one detection method can probe the full parameter range of the exoplanet population \citep{Gaudi2021}. 

While statistical analysis has estimated the observational bias in subsets of the data to calculate occurrence rates of different planet classes \citep[see review by][]{Zhu2021}, there is not a satisfactory way to work backwards from this to remove bias from a catalog of observed properties for machine learning. Steps such as limiting analysis to a subset of data considered complete is both difficult to ensure and would require abandoning a large fraction of the archive, restricting imputation of planet properties to only a small region of the parameter space. We therefore do not add any correction to the dataset, and treat the imputed planet properties as ``synthetic observations'' with errors similar to the measured values for that planet. These are used to explore the demographics of the currently observed planet population, which must form the ground truth for any theories for planet formation.

\section{Method and algorithms}
\label{sec:method}

The five numerical data imputation methods that are compared for imputing missing values in the exoplanet archive are the $k$NN-Imputer, MissForest, GAIN, MICE, and the newly proposed \kNNKDE. All five methods have the ability to utilize an incomplete dataset with missing values. In section~\ref{sec:completedata}, where a dataset of complete properties is used as the first benchmark, performance is also compared with the modified Boltzmann Machine (mBM) neural network presented in TLG2020. In addition to imputing a single value for a missing entry, the \kNNKDE developed for this work capable of providing a probability distribution for the imputed value. Below, each of the numerical imputation algorithms are briefly described, along with hyperparameter values. 

\textbf{The $k$NN-Imputer} uses the traditional $k$-nearest neighbors algorithm to fill-in missing observational values \citep{Troyanskaya2001}. For each observation (planet) in the dataset with missing properties, the algorithm computes the distance to every other observation using the Euclidean distance between properties that are observed in both cases. The missing property is then imputed using the average of the $k$ nearest neighbors that have that value observed. The latter step results in different neighbors being used to impute different properties, which can potentially lead to inconsistent values (e.g. an imputed planet mass and radius which lead to an unphysical density). Despite its simplicity, the $k$NN-Imputer has been shown to provide robust and accurate numerical imputation results \citep{Emmanuel2021, Jager2021, LalandeDoya2022}. The hyperparameter $k$ which controls the number of neighbors can be optimized, and the value is fixed to $k=15$ throughout this study.

\textbf{MissForest} is an iterative imputation algorithm which uses Random Forests for regression \citep{Stekhoven2012}. All missing values are initially filled with the column mean (planet property mean). MissForest then considers each property individually, and replaces the initial mean values with imputed values based on a Random Forest regression using all other properties. After performing this step for each property once, the process can be repeated again. Observed values always remain unchanged, while the missing estimates are updated. MissForest stops either after a fixed number of iterations have been performed or when the imputed values have sufficiently converged. This numerical imputation method can be computationally expensive, but has shown great practical results and is flexible to heterogeneous dataset types and structures \citep{Lin2020, Jager2021, LalandeDoya2022, Grinsztajn2022}. The hyperparameter corresponding to the number of trees employed by the Random Forest algorithm is set to $N_{\rm trees}=20$.

\textbf{GAIN} (Generative Adversarial Imputation Nets) is an artificial neural network which revisits the GAN (Generative Adversarial Network) framework to impute missing values in numerical datasets \citep{Yoon2018}. Standard GAN models are composed of a generator and a discriminator. While the generator is tasked to output realistic observations, the discriminator aims at differentiating between real observations (from the dataset) and fake observations (synthesized by the generator). Unlike standard GAN models which generate entire observations, GAIN works on a cell by cell basis with the intent to fill-in missing values with credible synthetic data. This method claims state-of-the-art numerical data imputation results and has benefited from a lot of attention recently. However, recent benchmarks indicate that GAIN practical performances are mediocre when employed with real data sets \citep{Jager2021, LalandeDoya2022}. GAIN additionally has many hyperparameters to tune, such as the batch size, the hint rate (fraction of correct labels to hint at the discriminator), the number of training iterations, and an additional weight parameter $\alpha$ used to help the generator to mimic original observed data. In this work, only the number of training iterations was optimized, which is the the most sensitive hyperparameter for GAN models. A value of $N_{\rm iter.}=2,500$ was selected for best performance. 

\textbf{MICE} (Multiple Imputation Chained Equations) is another iterative imputation algorithm which uses linear regression \citep{VanBuuren2011}. Similar to MissForest, missing values are first filled with the column mean (i.e. the mean of the planet property), and the algorithm then loops over every column, one at a time. For a given column, MICE employs a linear regression to estimate missing values, unlike MissForest which uses Random Forests. After the linear regression, missing values estimates are updated while original values remained untouched. MICE stops either after a fixed number of iterations has been performed or when the missing values estimates have sufficiently converged. This algorithm is appreciated for its simplicity, its low computational cost, as well as its absence of any hyperparameters. However, it is not able to capture non-linear dependencies. 

Finally, \textbf{the \kNNKDE} is a newly proposed imputation algorithm, inspired by the traditional $k$NN-Imputer, and specifically tailored to return probability distributions for each missing value in an incomplete dataset \citep{Lalande2023}. As with the $k$NN-Imputer, the \kNNKDE searches for neighbors to a planet with missing properties by considering the Euclidean distance between those properties that are observed. However, in contrast to the $k$NN-Imputer, the \kNNKDE looks only for neighbors which have observed values for all missing properties to be imputed, rather than considering the properties individually. This ensures that imputed values are consistent with one another. Probability distributions are modeled as a mixture of Gaussians via Kernel Density Estimates (KDE) where each neighbor value is additionally weighted by distance to give greater importance to planets whose known properties are in close agreement. The hyperparameter of the \kNNKDE is fixed to $\tau = 50^{-1}$ for the softmax temperature, which controls the tightness of the effective neighborhood around each observation. A higher value for $\tau$ would distribute weights more uniformly across all neighbors, while a lower value would distribute most of the weight to the nearest neighbor. In addition, a new hyperparameter is introduced for the \kNNKDE, which limits the total number of neighbors considered for imputation. Its value is fixed to $N_{\rm cap} = 20$ throughout this study. Capping the total number of neighbors prevents the algorithm from using distant and irrelevant observations, potentially leading to broad average in dense areas of the parameter space (e.g. hot Jupiters or super Earths). As the \kNNKDE returns probability distributions (and not point estimates), estimating missing values is done by computing the mean of the distribution when needed.

By default, the \kNNKDE will impute all missing properties for a planet. However, the algorithm can also impute a subset of missing values. This allows properties with a high missing rate to be included in the dataset to guide the imputation, but not be imputed themselves. Such a step was introduced for section~\ref{sec:extended}, where the high missing rate of the additional properties included in the dataset would force the neighbor search to extend over a unhelpfully large region to find measured values.

\subsection{Including the minimum mass}
\label{sec:method_minmass}

After the transit technique, the most successful planet detection method is the radial velocity technique. A planet detected via the radial velocity method will have a measured minimum mass due to the unknown inclination of the planet orbit. The minimum mass is an important proxy for estimating the true mass, but as it is a lower limit on the actual value, it cannot be passed to the algorithms as one of the measured planet properties. For the algorithms that return a single value, a minimum mass observation can therefore not be included when imputing missing values. However, the additional information from the minimum mass can be leveraged with the \kNNKDE by computing the convolution of the estimated probability distributions with the distribution of possible masses based on the minimum mass observation.

In section~\ref{sec:impute}, the performance of the \kNNKDE is tested on synthetic radial velocity observations, where both the measured planet radius and mass are concealed from the algorithm and a minimum mass measurement is generated. Following the same methodology as TLG2020, $100$ minimum masses are generated for each planet with a concealed mass and radius value by randomly drawing orbital inclination values following a sinusoidal distribution. The convolution between the estimated mass distribution from the \kNNKDE algorithm and the distribution of possible masses given an observed minimum mass is performed as described in TLG2020. In practice, this corresponds to computing and assigning new weights to the individual mass-radius bivariate samples returned by the \kNNKDE.

To give the same bulk properties as for a transit observation, we show the imputed results by the \kNNKDE for both planet mass and radius for the radial velocity observation tests. The final mass and radius estimates by the \kNNKDE are computed by taking the average of the distribution after convolution with the minimum mass. Because this procedure is repeated for $100$ different possible minimum mass measurements, each planet has $100$ pairs of mass and radius estimates. The final estimate is therefore taken as the mean over the $100$ estimates. Note that the quoted error for mBM in section~\ref{sec:cd_rv} differs from that in TLG2020 due to a small difference in the definition of the overall error that better matches with the plotted data. Instead of taking the average over the 150 planets and the 100 convolutions in a single step, the average over the 100 convolution is performed first to provide the mass and radius estimates. The RMSE is then computed over the 150 planets in the test set. This way of computing the error more faithfully reflects the data, particularly the visual data plotted in Figure~\ref{fig:cd_rv_mass_radius} which already represent the average taken after 100 convolutions.

Of course, a real observation would only include a single minimum mass measurement. The resulting error might therefore be significantly greater than the reported average, depending on the degree of inclination of the planet orbit.

\section{Imputing planet mass}
\label{sec:impute}

The five machine learning algorithms presented in section~\ref{sec:method} draw on different techniques to impute missing values using the large but incomplete exoplanet archive. The advantage over previous codes is the ability to utilize all planet discoveries, rather than basing the imputation on a small subset of the known population. But it first needs to be confirmed whether the addition of incomplete data can assist the accuracy of the imputation.

In section~\ref{sec:completedata}, the performance of the algorithms is therefore initially tested on a complete dataset with no missing values. This {\it complete properties dataset} is the same that was used in TLG2020 and consists of 550 planets with observed values for six properties: planet radius, planet mass, orbital period, planet equilibrium temperature, stellar mass and the number of known planets in the system. The performance of the algorithms using this dataset is compared with the mBM neural network in TLG2020 that requires complete data, and forms a baseline comparison for the next two datasets.

The {\it full archive dataset} that is used in section~\ref{sec:fullarchive} includes the same six properties, but no longer requires every planet in the dataset to have a complete set of observed values. This greatly expands the dataset size to 5,251 planets (the full set of discovered planets listed by the NASA Exoplanet Archive on February 2, 2023, including Solar System planets). Comparison with the results from the previous complete properties dataset assesses the value of using large but incomplete data.

The {\it extended dataset} that is used in section~\ref{sec:extended} contains the same 5,251 planets, but now adds two additional properties to be used in the imputation: stellar metallicity and planet orbital eccentricity. This comparison looks at the value of additional information for the mass imputation.

For each of the three datasets, the algorithms are tested by artificially concealing known planet mass values and then imputing the mass with each code. To replicate two of the most likely use cases, we perform two tests for the mass imputation. The first test conceals only the planet mass, and the algorithms are tasked to impute the mass based on all other observed properties. In the case of the complete properties dataset in section~\ref{sec:completedata}, this replicates an observation performed with the transit technique, which is the planet detection method employed by the dedicated space-based planet hunting missions and accounts for over around 75\% of the planet discoveries. The transit technique measures a planet radius, but not planet mass. All planets in the complete properties dataset have a measurement for the planet radius. For the full archive and extended datasets, the planet radius is used when available and an error value is computed for the same planets present in the complete properties dataset (with radii values) for accurate comparison.

The second test of the mass imputation resembles an observation using the radial velocity method, which provides a minimum mass for the observed planet, $m_m = m_p\sin(i)$ for orbital inclination, $i$, and no radius measurement. Both planet mass and planet radius are therefore initially concealed and imputed by the algorithm. The minimum mass observation is then included by convolving the probability distribution of possible masses and radii values with the univariate distribution of possible masses from the minimum mass measurement (see section~\ref{sec:method_minmass}). This second test was only possible for the \kNNKDE algorithm and the previous mBM code, which can produce distributions (rather than point estimates) for imputed properties.

\subsection{The complete properties dataset}
\label{sec:completedata}

The performance of the five algorithms was first tested on the 550 planet dataset used in TLG2020. Each planet entry in that dataset had six measured properties for planet radius, planet mass, orbital period, planet equilibrium temperature, stellar mass and the number of known planets in the system, with no missing values. To compare directly with TLG2020, 150 planets within that dataset were assigned as a test set, with selected properties artificially hidden and imputed by each algorithm. The same 150 planet test subset that was presented in the main results section in TLG2020 is used to test the five proposed algorithms.

As the new algorithms can leverage incomplete data to estimate missing values, all 550 planets in this dataset were provided to each imputation method (where 150 planets had one or two missing property values) to estimate the missing properties. This differs from TLG2020, where the mBM network needed to be trained on the 400 planets with complete properties, and the resulting relative probability density function created by the mBM was then used to impute the missing properties in the test set. 

\subsubsection{Mass prediction in the transit regime: complete properties dataset}
\label{sec:cd_transit}

\begin{figure*}[!ht]
    \centering
    \includegraphics[width=0.95\textwidth]{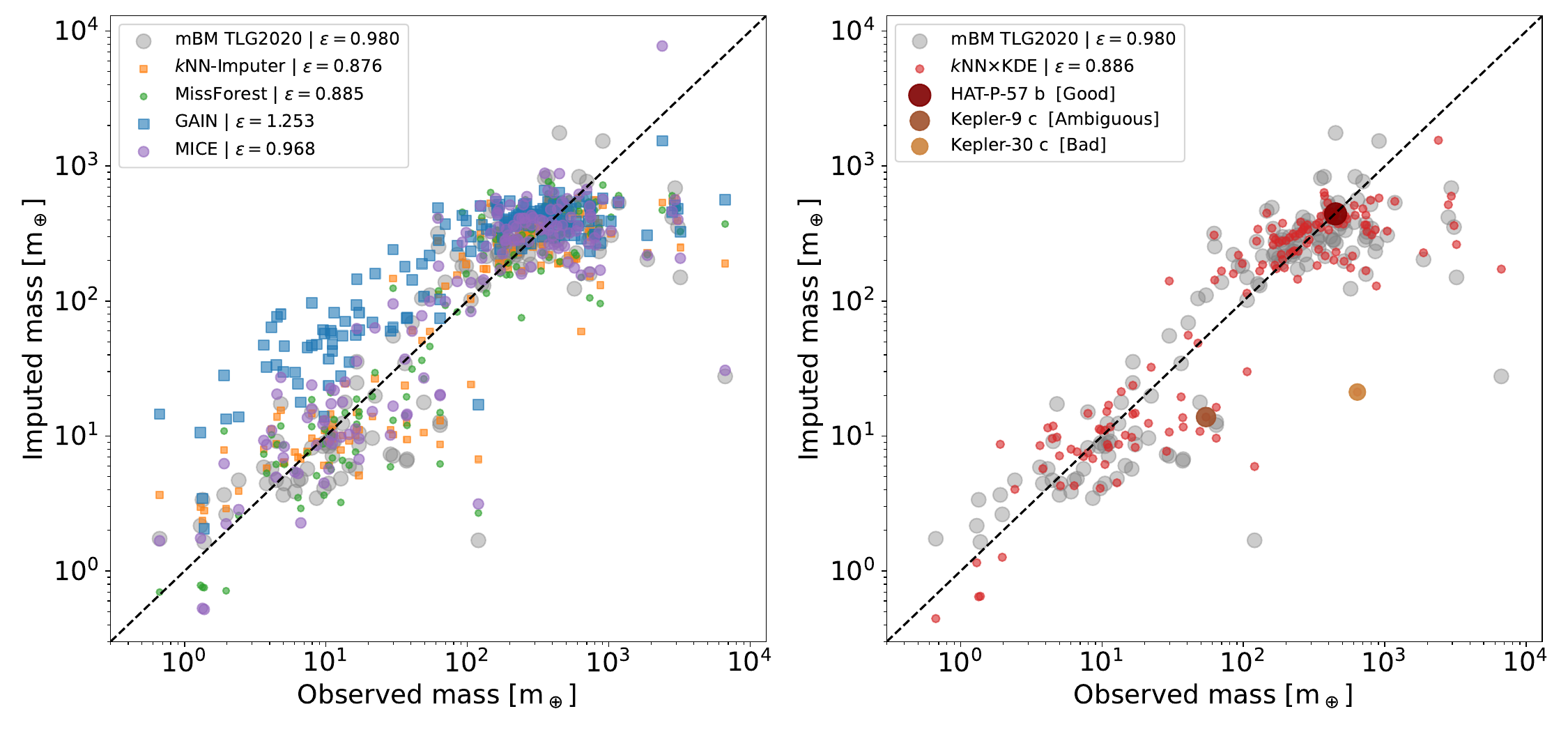}
    \caption{Test results when using the complete properties dataset where the 150 test planets are treated as transit observations, with missing mass values. The left-hand plot shows four proposed imputation algorithms alongside the mBM code in TLG2020. The right-hand plot shows the comparison between the observed mass and imputed mass for the mBM code and the \kNNKDE algorithm. The figure legend shows the average error across all 150 plotted planets. The diagonal dashed line marks a perfect correspondence between the observed and imputed values. The distributions of the three planets marked in the right-hand legend are shown below.}
    \label{fig:cd_transit_mass}
\end{figure*}

Figure~\ref{fig:cd_transit_mass} shows the imputed mass compared to the observed mass value for the 150 test planets when their observed mass value was concealed and imputed by each algorithm. This simulates the performance for estimating mass from a transit detection, which can provide five of the six dataset properties, but no mass measurement. The left-hand plot of Figure~\ref{fig:cd_transit_mass} shows the performance of four of the new codes compared with the mBM of TLG2020 shown in grey. On the right-hand plot, the \kNNKDE algorithm marked in red is compared with the mBM. 

For each planet, $p$, that has an imputed mass, the error is computed by taking the natural logarithm of the ratio of the observed planet mass ($m_{\rm p,o}$) and the imputed mass ($m_{\rm p,i}$) to give $\epsilon_{\rm p} = \ln\left(m_{\rm p,o}/m_{\rm p,i}\right)$. These values are averaged over the 150 test planets, $N$, by taking the root mean square to give the final reported error for each algorithm as $\epsilon =\sqrt{\left(\sum_{\rm p=1}^{N} \epsilon_{\rm p}^2\right)/N}$. This error value is shown in the top left corner of Figure~\ref{fig:cd_transit_mass} for each of the five algorithms and the mBM. A perfect match, with an error of $\epsilon_{\rm p} = 0.0$, would lie along the diagonal dashed line. 

A comparison of the average error from all five codes and the mBM shows that four out of the five new imputation techniques surpass the original result, indicating that the algorithms in this paper are competitive with the mBM in addition to the flexibility to use on incomplete data. Two main groups of planets can be seen in the distribution of planet masses, one in the gas giant regime (with masses similar to that of Jupiter) and the second in the smaller super-Earth regime, with masses about ten times that of the Earth. Planets do exist between and outside these two groups, but are less clustered. All algorithms perform best in the mass regime of Jovian planets. This is not surprising, as large gas giants are typically easier to detect than small rocky worlds, providing a more densely packed parameter space over that mass range.

The poorest performance was that by the GAIN algorithm, which can visually be seen as an average overestimate of the planet mass over the full range of masses in the dataset. In spite of impressive results provided by deep-learning models for image, video, or text data, it has been shown that statistical methods (and in particular tree-based models) remain the state-of-the-art for numerical and tabular datasets \citep{Grinsztajn2022}. This state of affairs is observed here, where the $k$NN-Imputer and the \kNNKDE (nearest-neighbor methods), as well as MissForest (a tree-based model), providing with best results, with an error of around $\epsilon = 0.88$ that corresponds to a factor of 2.4 from the observed mass. Statistical tools have very few hyperparameters to train which not only prevents from overfitting (i.e. when the model predictions are accurate only for the training dataset), but also facilitates results interpretation. Conversely, the GAIN is a generative adversarial network which necessitates to fine-tune thousands of trainable parameters. Generative adversarial networks are known for being particularly hard to train, to interpret, and to diagnose \citep{Saxena2021}. Notably, they easily suffer from ``mode-collapse problem'', where the output distribution shrinks down to a small region of the desired target distribution. This is what is happening here: the dominating Jupiter planets mislead the GAIN into generating planets mostly in this regime, therefore over estimating the mass of planets in the Super-Earth regime.

\begin{figure}
    \centering
    \includegraphics[width=0.99\columnwidth]{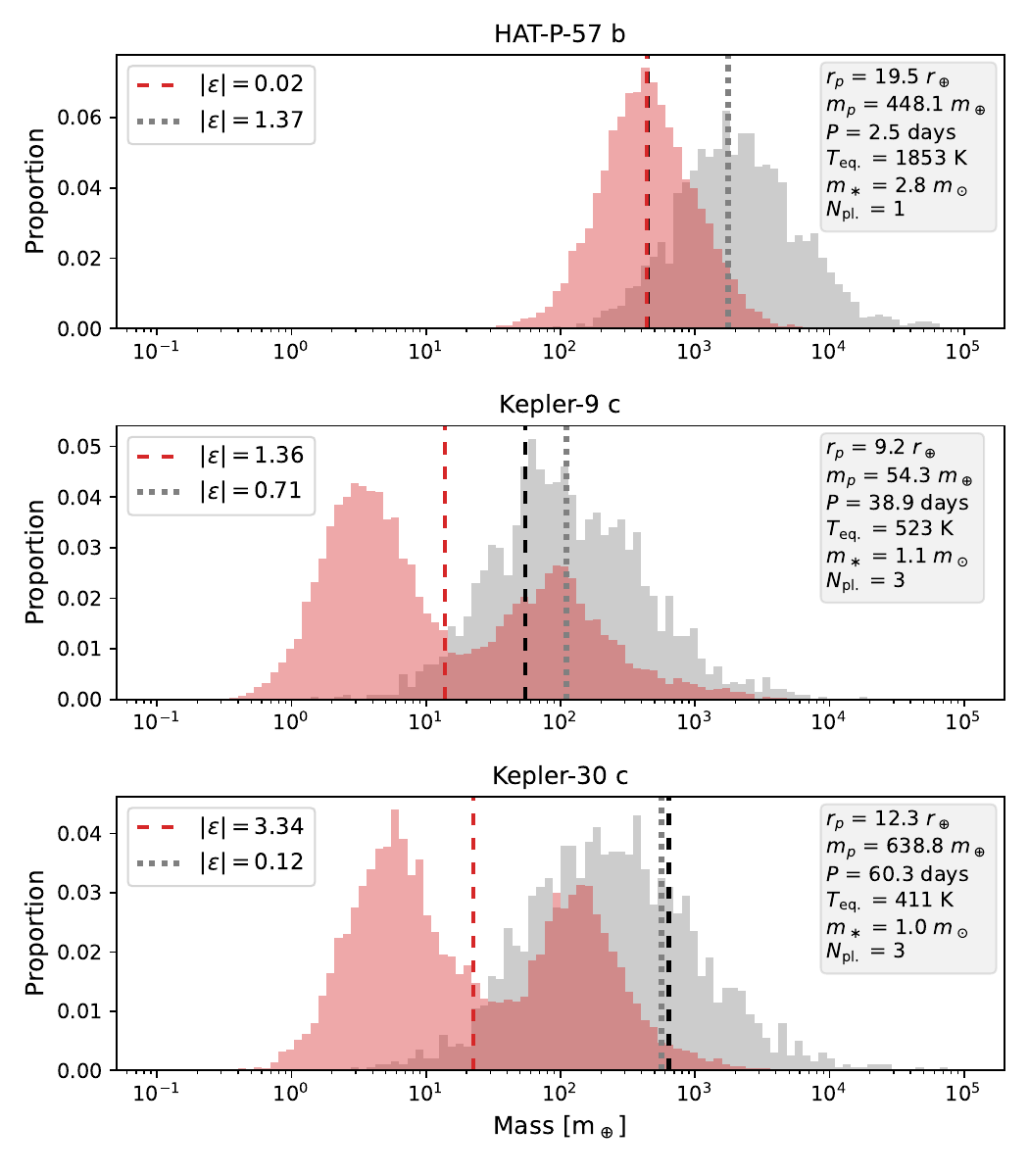}
    \caption{Distributions of the imputed mass values calculated with the \kNNKDE for the three planets highlighted in Figure~\ref{fig:cd_transit_mass}, selected for their low error (top) and high errors (middle and bottom). The red histogram shows the distribution calculated by the \kNNKDE, and the gray histogram is the distribution from the mBM in TLG2020. The vertical black line is the observed mass for the planet, while the red and gray vertical lines show the imputed value from the \kNNKDE and the mBM, respectively. Top panel shows the mass distribution for HAT-P-57b: a hot Jupiter with a low error for the imputed mass. The middle and lower panes show the mass distributions for Kepler-30c and Kepler-9c, both of which have higher errors due to being unusual planets within the dataset.}
    \label{fig:cd_transit_profiles}
\end{figure}

\subsubsection{Mass distributions in the transit regime: complete properties dataset}
\label{sec:cd_transit_distrib}

In addition to a single imputed mass value, the \kNNKDE algorithm can provide a probability distribution for the imputed value thanks to weighted kernel density estimates described in section~\ref{sec:method}. This can be used to understand the origin of the estimated value, which can reveal information about the underlying demographics of the observed planet population. To assess the value of this additional feature, the mass distributions for three planets imputed by the \kNNKDE (red histogram) and the previous distributions from the mBM in TLG2020 (gray histogram) are shown in Figure~\ref{fig:cd_transit_profiles}. These are the same three planets indicated by enlarged circles on the right panel of Figure~\ref{fig:cd_transit_mass} and were selected for their particularly low and high error value.

\paragraph{(Low error) HAT-P-57b: A typical class of planet} The top-most pane in Figure~\ref{fig:cd_transit_profiles} shows the mass distribution for HAT-P-57b, which has one of the lowest errors in the dataset. HAT-P-57b is a hot Jupiter, with a mass of 1.41\,\MJ, radius of 1.74\,\RJ and orbital period of just 2.5 days \citep{Hartman2015, Stassun2017}. Hot Jupiters were among the first extrasolar planets to be discovered, as both the radial velocity and transit techniques are most sensitive to planets with large sizes and short periods. Because of this observational bias, hot Jupiters are well represented in the exoplanet archive, despite only orbiting about 1\,\% of stars \citep{Wright2012}. This is particularly true of this complete dataset, as the requirement to have both planetary mass and radius measurement usually requires both a transit and radial velocity detection, bumping the occurrence rate of the more easily detected Jovian-sized planets (M$_p > 0.1\,$\MJ) within the database to 72\,\%, 86\,\% of which have orbits shorter than 10 days.

As a result, HAT-P-57b sits in a dense area of the parameter space for all six of the planet's observed properties (this can be seen visually by estimating the planet's position on Figure~\ref{fig:pairplot}) and the imputed values for hot Jupiters typically have the lowest errors across all the algorithms. The relatively narrow profile width indicates that the 20 nearest neighbors selected by the \kNNKDE in the six-dimensional property space to be used for imputation all have similar masses. This suggests HAT-P-57b belongs to a typical class of known exoplanets, and the algorithm is confident in favoring a Jovian mass world based on the five provided measurements. The previous estimate by the mBM for TLG2020 also predicted that HAT-P-57b would be Jovian size, but estimated a larger mass with a broader distribution of possible values.

\paragraph{(High error) Kepler-9c \& Kepler-30c: Peak selection} The middle and lower panes of Figure~\ref{fig:cd_transit_mass} show the mass distribution for planets with particularly high errors, with Kepler-30c (lower pane) having one of the largest errors in the test set. In both cases, the reason for the high error is that the imputed value is an average between two possibilities for the mass. Based on the five observed properties available to the algorithm, the \kNNKDE therefore considers that Kepler-9c and Kepler-30c could be either gas giants with a mass around 0.5\,\MJ, or super Earths with a mass around 3\,\ME.

For the \kNNKDE, this dichotomy highlights one of two situations. Either the five measured properties of Kepler-9c and Kepler-30c can belong to two different mass regimes of planets, or the pair are unusual within the exoplanet demographic and their twenty neighbors are covering a wide area of the parameter space. In this case, a comparison with the planets' observed properties listed in Figure~\ref{fig:cd_transit_profiles} with the demographics pairplot in Figure~\ref{fig:pairplot} reveals that it is the second option: both planets are a little unusual.  

The size of Kepler-9c is rare within current exoplanet discoveries, consistent with a inflated sub-Saturn \citep{Holman2010, Borsato2019}. This places Kepler-9c between the two most common sizes for discovered extrasolar planets, sitting at the dip in the radius histogram in Figure~\ref{fig:pairplot}, and the thin neck of the planet radius versus planet mass plot. This parameter region is particularly sparse for the complete dataset used here (black dots in Figure~\ref{fig:pairplot}) where very few examples of planets exist with radii between the super Earth and gas giant regime. It is therefore not surprising that the \kNNKDE algorithm has found that the closest matches the planet's known properties have masses both higher and lower than the ground truth. 

Despite being aware of more larger radii planets than small, the \kNNKDE favors the smaller super Earth mass peak for Kepler-9c. This is probably because Kepler-9c resides in a system with three known planets. Multi-planet systems have been found to favor similar sized worlds with smaller sizes \citep{Weiss2018}, reducing the probability that such planets are gas giants. The nearest neighbors to Kepler-9c are therefore more likely to have lower masses. 

One of the main advantages of a returned distribution is the ability not to settle for the returned imputed value. In the case of a multi-modal profile, it does not make sense to select the average value as the imputed mass, but rather select one of the peaks. Given the large radius for Kepler-9c and the reasonably similar size of both mass peaks, a researcher wishing to estimate the mass value might select the higher mass peak as the most probable value, but note that the presence of the second peak meant the planet's size in a multi-planet system was rare. In this case, the resulting estimation would be close to the observed value. 

The origin of the bimodal profile for Kepler-30c is more intriguing. Unlike Kepler-9c, the planet's observed radius is not ambiguous: at 1\,\RJ, Kepler-30c is a gas giant, although its observed mass at 2\,\MJ is at the upper end for planets of that size \citep{SanchisOjeda2012}. Given the certainty for HAT-P-57 b, it therefore seems initially surprising that the resulting mass profile do not strongly peaks at a gas giant mass. However, the orbital period for the planet at 60 days is quite long and past the typical distance for hot Jupiters, which usually have orbital periods less than about 10 days. As can be seen from the Figure~\ref{fig:pairplot}, this 60 day orbit makes the planet a more unusual find, and the surrounding parameter space in that property is relatively sparse, with the majority of planets at that distance having super Earth masses.

Despite the unusual period, there are sufficient planets close to Kepler-30c in the visible five properties for the \kNNKDE algorithm to create a relatively close group in that parameter space with which to estimate the mass. But the mass for those neighbors turns out to cover a large range than for the other properties, resulting in the bimodal profile with the peak at super Earth masses. This indicates that planets on these longer more temperate orbits have a surprisingly wide range in density. It is an interesting trend to note, and potentially might suggest that this group of planets have a range of compositions driven by their evolution, which might include migration from the outer, icy parts of the protoplanetary disk, or in-situ formation. However, this might also be due to the challenges of detecting planets at these distances from the host star. Many planets in multi-planet systems on longer orbital periods like Kepler-30c have their mass measured via transit timing variations (TTV). TTV can often gives quite high errors on the mass, which may explain the wide range in densities amongst Kepler-30c's neighbors, rather than strong variations in their mineralogy. 

The true observed mass for Kepler-30c is at the highest end of the imputed distribution. There is a definite bump around that location, but it is not considered the most probable result. In this case, manually selecting a peak is not enough to get an accurate imputation, but the presence of multiple peaks is still a flag to investigate the origin of the imputation which reveals the uncertain demographics of the surrounding population. 

Interestingly, the previous mBM model is broader and not bimodal for either Kepler-9c or Kepler-30c, considering both planets likely to be gas giants. This may be because the neural network has internally weighted the importance of radius measurement in indicating mass more highly than other properties. In these cases, that produces a more accurate answer, but the neural network is more opaque than the statistical \kNNKDE, so reveals less about the underlying planet demographics.

\subsubsection{Mass and radius prediction in the RV regime: complete properties dataset}
\label{sec:cd_rv}

The second test using the six complete properties dataset simultaneously conceals and imputes both the mass and radius values, and then weights the resulting distributions with a minimum mass measurement as described in section~\ref{sec:method_minmass}. This replicates imputing mass and radius values for a radial velocity observation, where a minimum mass, orbital period, effective temperature, stellar mass, and number of known planets would be expected data, but no radius or true mass measurement. Since the convolution step with the minimum mass is only possible where a distribution of values can be imputed, only the \kNNKDE algorithm can be compared with the previous mBM neural network for this test. 

\begin{figure*}[!ht]
    \centering
    \includegraphics[width=0.95\textwidth]{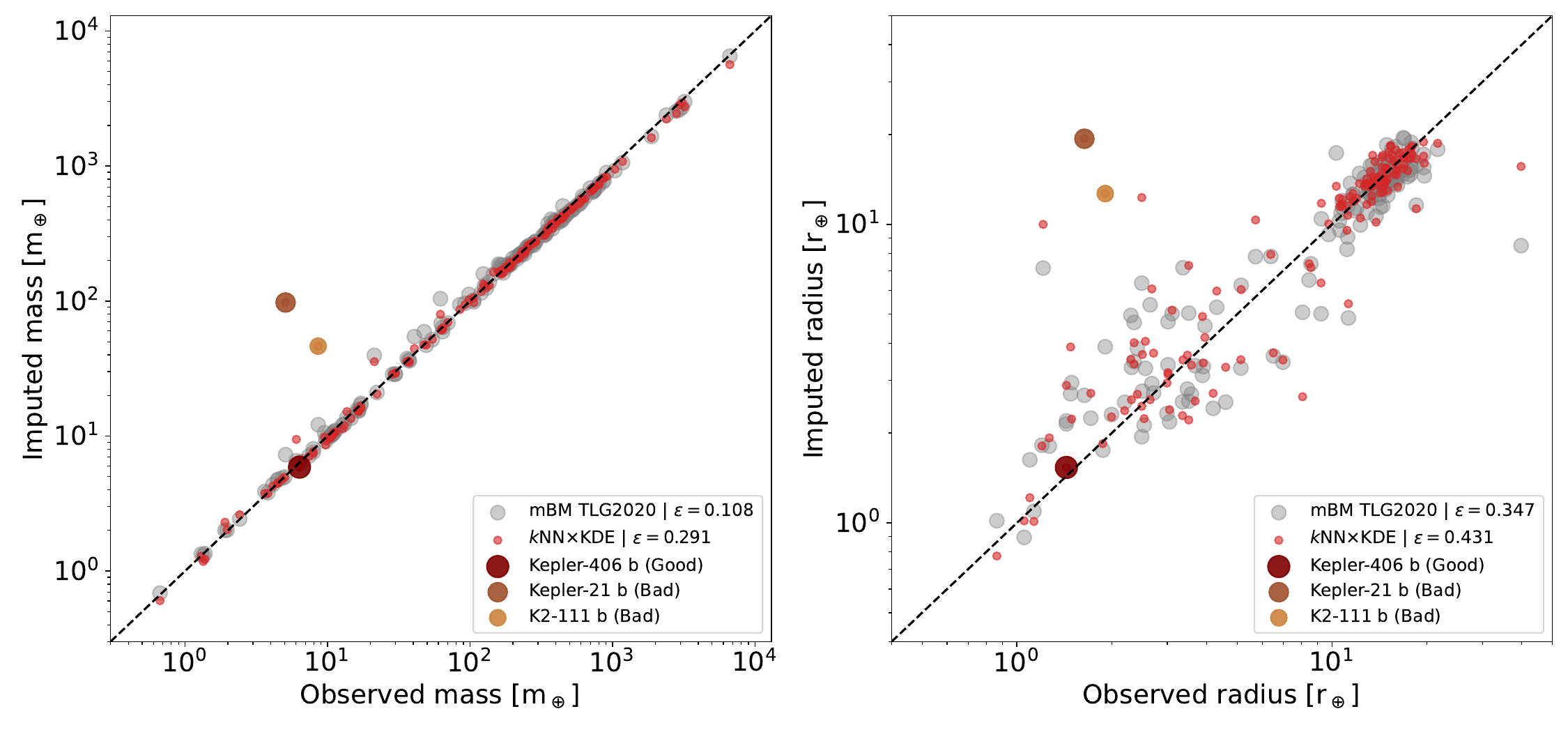}
    \caption{Test results when using the complete data set where the 150 test planets are treated as radial velocity observations, with missing radii values and a minimum mass measurement. The left-hand plot shows the results for the mass imputation, after the imputed distribution has been weighted by the minimum mass. Red dots show the \kNNKDE algorithm results, while the light grey distribution is from the mBM code in TLG2020. The right-hand plot shows the comparison between the observed radius and imputed radius values that correspond to the weighted imputed masses. The mass and radii imputed value distributions for the three highlighted planets are shown in the next figure.}
    \label{fig:cd_rv_mass_radius}
\end{figure*}

The left-hand plot in Figure~\ref{fig:cd_rv_mass_radius} shows the imputed mass versus the observed mass for the same 150 test set planets as in section~\ref{sec:cd_transit}, where this time both the mass and radius values have initially been concealed before weighting with a minimum mass value. The low error and tightness of the fit compared to the transit regime test in Figure~\ref{fig:cd_transit_mass} is a reflection of the importance of the minimum mass measurement compared to radius in imputing the planet mass. This can be compared to the right-hand plot of Figure~\ref{fig:cd_rv_mass_radius}, which shows the radius imputation where the minimum mass measurement has been used. The scatter here is visually similar to Figure~\ref{fig:cd_transit_mass}, where the mass is being imputed with radius information.

\begin{figure*}[!ht]
    \centering
    \includegraphics[width=0.85\textwidth]{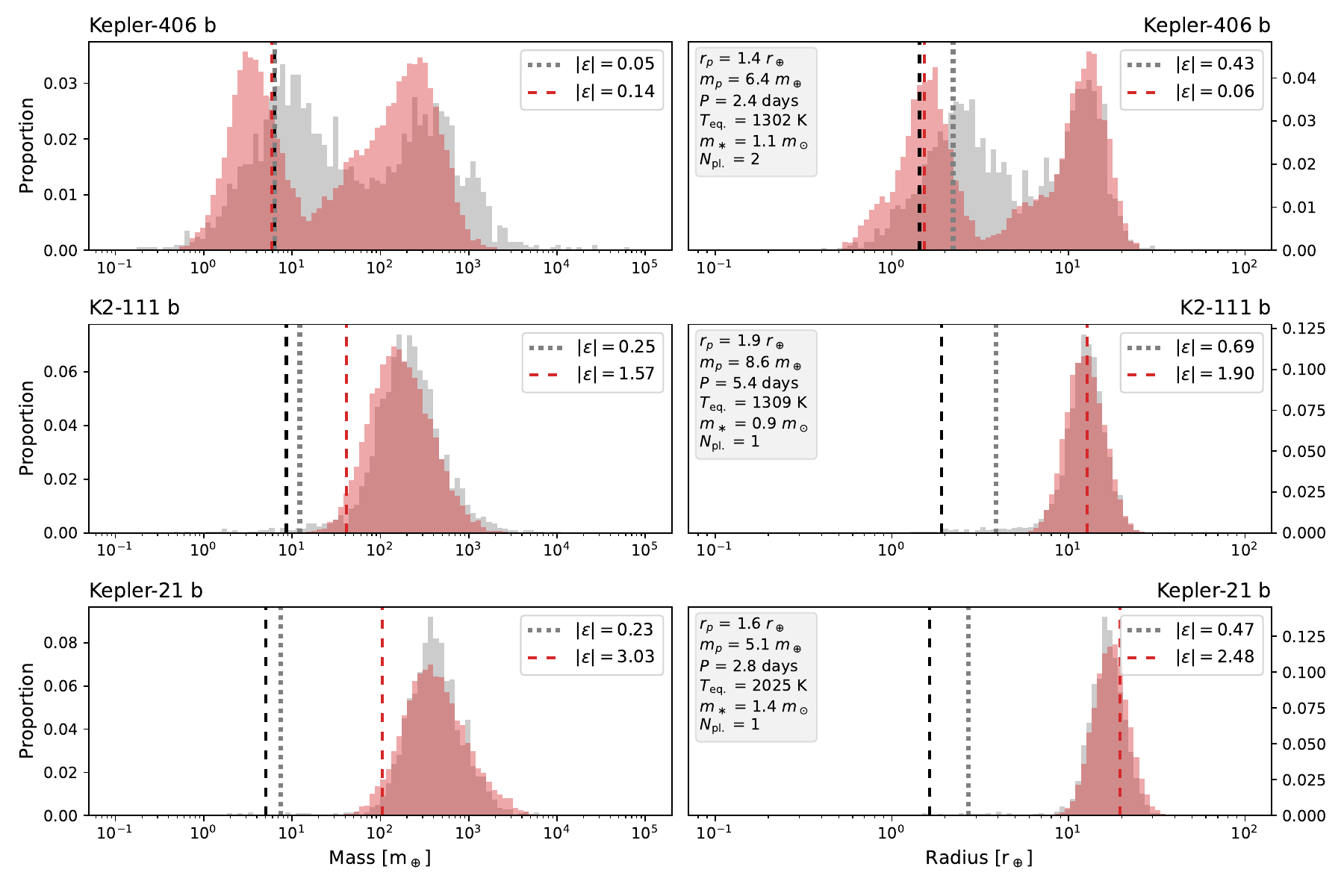}
    \caption{Distributions of the imputed mass and radius values calculated with the \kNNKDE for the three planets highlighted in Figure~\ref{fig:cd_rv_mass_radius}, chosen for their low error (top) and high errors (middle, bottom). The red histogram shows the distribution calculated by \kNNKDE, while the gray histogram is the distribution from the mBM in TLG2020. The distributions are from the initial imputation of the algorithm, before the weighting with the minimum mass distribution. The vertical black line is the observed mass for the planet, while the red and gray vertical lines show the imputed value from the \kNNKDE and mBM, respectively. The top panels show the distribution for Kepler-406b, which has a very low error. The next two distributions fro K2-111b and Kepler-21b both have particularly high errors.}
    \label{fig:cd_rv_profiles}
\end{figure*}

The average error, $\epsilon$, for the \kNNKDE is higher than the mBM in TLG2020. The error is exacerbated by particularly poor imputations for the mass and radii of Kepler-21b and K2-111b, which are highlighted in larger circles in Figure~\ref{fig:cd_rv_mass_radius}, and whose mass and radii distributions imputed by the codes (prior to weighting by the minimum mass distribution) are shown in the second and third panel of Figure~\ref{fig:cd_rv_profiles}. From the distributions, it is evident that both \kNNKDE and the mBM believe that these two planets should have a higher mass and radius than observed. In fact, the two codes strongly agree with one another that the expected masses and radii are those of a Jovian-sized gas giant. This indicates that there is something unusual about finding a planet of smaller size in these particular environments.

\paragraph{(High error) K2-111b \& (low error) Kepler-406b: A missing observation} The unusual nature of K2-111b can be understood by comparing the distribution to that of a very low error planet, Kepler-406b (top pane in Figure~\ref{fig:cd_rv_profiles}) \citep{Marcy2014}. Unlike the distributions for K2-111b or Kepler-21b, both \kNNKDE and mBM propose two almost equally possible options for the mass and radius of Kepler-406b: one Jovian-sized planet and one super-Earth possibility. With neither a planet mass nor radius measurement available to the codes, the imputation is based on the planet's orbital period, equilibrium temperature, stellar mass and number of known planets in that system. In three of these properties, both the low error Kepler-406b and the high error K2-111b are very similar. However, Kepler-406b was known to be in a two-planet system, whereas K2-111b was thought to be alone \citep{Fridlund2017}. As was discussed in section~\ref{sec:data}, the bottom row on the pairplot in Figure~\ref{fig:pairplot} shows a trend towards massive and closely orbiting planets in single planet systems where a migrating hot Jupiter may have suppressed additional planet formation, but a more even range of masses where two planets are present. The appearance of the second, lower-mass peak for Kepler-406b was therefore driven by the inclusion of the second system planet. In this situation, the minimum mass measurement resolves the dichotomy, causing the algorithm to select the lower mass peak as the most likely value.  

These two distributions for Kepler-406b and K2-111b indicate that both super Earths and gas giants are commonly found on orbits of a few days around solar-type stars. However, in the case where only one planet orbits, a hot Jupiter whose migration is capable of disrupting further planet formation is more common. This then leaves the question as to why K2-111b is not a gas giant, but has a measured mass of a super Earth. The answer is that the data for the K2-111 system is actually incomplete. The dataset used in this section was the same dataset for TLG2020, and taken from the exoplanet archive in 2018. In 2020, a second planet was discovered orbiting K2-111 \citep{Mortier2020}. The presence of this second planet was therefore hinted at by the high error of the \kNNKDE and mBM result based on the demographics of the planets in the dataset, as the mass of K2-111b would be more commonly found in a multi-planet system. This demonstrates that numerical imputation schemes that can return distributions such as the \kNNKDE can be used not only to impute values, but also to investigate known planet properties to see how typical they are amongst the known exoplanet population. It is information that could be useful for plans for follow-up discovery missions such as the proposed \textit{Japan Astrometry Satellite Mission for INfrared Exploration} (JASMINE), which will search for undiscovered planets in known systems \citep{Kawata2023}.   

\paragraph{(High error) Kepler-21b: An unusual size ratio} Both the \kNNKDE and the mBM also mistook Kepler-21b for a gas giant, rather than the observed super Earth. It is possible that this system also has a second, undiscovered planet. But in this case, the planet also has other  properties that also make its small size unusual. Kepler-21b closely orbits a very bright F-type star, giving the planet's properties a particularly high stellar mass and equilibrium temperature within the current demographics \citep{Howell2012}. Looking at the pairplot in Figure~\ref{fig:pairplot}, a trend exists between stellar mass and planets with large radii. There is also a weaker hint of a trend between planet mass and equilibrium temperature, with hotter planets typically being more massive. This results in the planets with closest matches to stellar mass and equilibrium temperature (neighbors in the parameter space) to Kepler-21b being gas giants. As mentioned in section~\ref{sec:data}, this trend may be due to the difficulty in finding smaller planets around more massive stars. The situation is particularly marked for the complete properties dataset used in this section, which requires a transit observation for every planet to measure the planet radius; a technique with a strong bias towards smaller ratios for the star to planet radius. It therefore seems likely that even a second planet in the system would not be sufficient to make Kepler-21b ``normal" within the exoplanet population and topple the likelihood that Kepler-21b is a gas giant. Instead, the high error reflects an unusual outlier in the exoplanet demographics, either due to observational constraints or a lack of examples for planets around F-type stars.

With the two profiles strongly overlapping, it is initially surprising that the mBM network does not also have a high error for these two planets. However, the mBM shows heavy-tailed distributions with a few outlying values at the lower end of the distribution. This tail artificially allows the convolution with the minimum mass measurement to favor an estimate close to the actual value, even when it lies in a very low probability region of the imputed distribution. In the case of the \kNNKDE distributions, the twenty neighbors all have high masses and radii, and a tail is not produced. The minimum mass therefore falls completely outside the range of estimated possible values. In this situation, the convolution with the minimum mass cannot sufficiently influence the final imputed value. The distance between the minimum mass value and the distribution peak can also indicate the high error for planets such as K2-111b and Kepler-21b even when the distribution shape looks reasonably certain. The minimum mass not lying within a high probability area of the distribution flags a disagreement between the two mass distributions, unless the orbit is especially inclined. This allows a similar investigation as above, even in cases where the true mass is not known. Conversely, a situation like Kepler-406b, where the minimum mass lies close to a mode of the mass distribution, would be expected to increase the likelihood of a low error.

\subsubsection{Summary: complete properties dataset}
\label{sec:cd_summary}

For this complete properties dataset, four out of the five new algorithms performed comparably on average with the mBM neural network, typically producing a slightly lower error result. This indicates that the ability to train on incomplete data does not degrade the performance where a complete set of properties is available. For the \kNNKDE where planet property distributions could be generated, the results frequently resembled the mBM, indicating that these two independent methods were extracting similar conclusions about the exoplanet demographics. In areas where the two differed, the \kNNKDE and mBM alternately achieved the better result. The origin of the \kNNKDE numerical algorithm is easier to understand in comparison with the pairplot in Figure~\ref{fig:pairplot}, but the mBM neural network can preferentially weight properties it deems more valuable. These are considerations when deciding what kind of code to use. 

In both the \kNNKDE and mBM case, the distributions of the imputed properties provide significantly more information than the imputed value alone. The distribution width can indicate the degree of certainty in the code, where a narrow profile suggests a more confident result. In the case of a multi-modal profile, selecting a peak rather than an average can improve the accuracy of the result. The shape itself indicates whether the currently observed planet demographics are consistent with multiple planet masses for the observed properties. When compared with a known observed mass value, the distribution can also highlight if the planet is an unusual outlier or even if an observed value is inconsistent with the other data. In the case of radial velocity observations, examining the distribution also distinguishes between an imputed planet size driven by the minimum mass observation, versus one where the imputed distribution from the code and that of the minimum mass have strong agreement. 

\subsection{The full archive dataset: six properties}
\label{sec:fullarchive}

The dataset used with the five algorithms is now extended from the 550 planet subset with six complete properties to using an incomplete dataset with missing values that includes all 5,251 planet discoveries. The factor ten increase in planet number also greatly increases the range in properties. While the complete properties dataset covered a mass range of about 1\,\ME to 5,000\,\ME, the full archive runs from approximately 0.1\,\ME to 10,000\,\ME (about 31\,\MJ: into brown dwarf territory). The mBM presented in TLG2020 is not able to train on incomplete data, so it is dropped from the comparison at this point. Without the necessity of a separate training set that was required by the mBM, the performance of the five algorithms is tested using a leave-one-out cross-validation method. This consists of removing one or more properties from one planet entry and imputing those missing values based on the remaining observations in the dataset. The process is repeated for each planet in the dataset. This method makes full use of the observed data, and prevents relying on a particular train/test data split.

As in the previous section, algorithm performance is tested in the ``transit regime" where known observed mass value are redacted and imputed to replicate the use case for a transit observation, and the ``radial velocity regime", when both mass and radius values are imputed with a minimum mass guide. 

\subsubsection{Mass prediction in the transit regime: full archive dataset} 
\label{sec:fd_transit}

\begin{figure*}[!ht]
    \centering
    \includegraphics[width=0.95\textwidth]{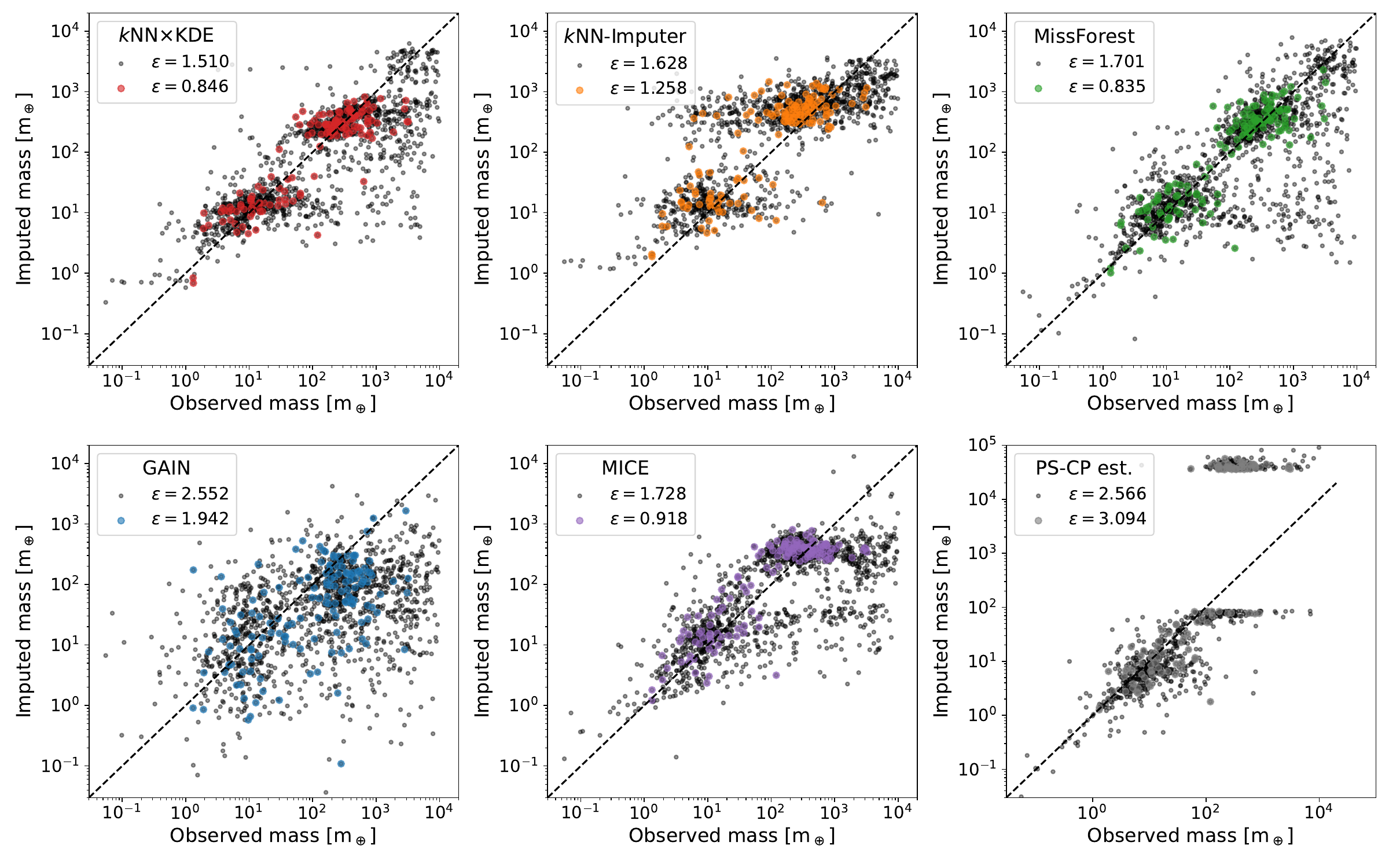}
    \caption{Test results when using the full exoplanet archive in the ``transit regime" where planet mass is concealed and imputed. Each of the five algorithms imputes the planet mass for planets with observed mass value, treating each planet as if it had been detected as a transit observation, with a radius observation recorded where available. The final pane shows the results of the mass–radius relationship of \citet{ChenKipping2017} used by the Planetary Systems Composite Parameters (PS-CP) table of the NASA Exoplanet Archive. The black dots in each panel show the entire database. The colored dots correspond to the planets in the same test set presented in section~\ref{sec:completedata} for comparative purposes. The two error values are for the complete dataset and the previous test subset.}
    \label{fig:fd_transit_mass}
\end{figure*}

Figure~\ref{fig:fd_transit_mass} compares the performance of the five algorithms in imputing planet mass. For each planet with a recorded mass observation (a total of 1,426 planets in the full archive dataset), the planet mass is concealed from the algorithm which then imputes the value based on the remaining observed properties for that planet and the proprieties of the other planets in the dataset. Note that while this test is referred to as a synthetic transit observation, the incomplete dataset means that it is no longer true that the imputed planet will always have a radius measurement. For comparison with the complete properties dataset, we therefore also calculate the error for the same 150 test planets with radius used in section~\ref{sec:completedata}.

Due to the number of planets now being tested, Figure~\ref{fig:fd_transit_mass} shows the results from each algorithm in a separate plot. Black dots show the imputed mass values for all planets tested, while the colored dots represent the planets that were also in the test set for the complete dataset in section~\ref{sec:completedata}. As it is no longer possible to compare with the mBM, the last panel in Figure~\ref{fig:fd_transit_mass} shows the results of the mass-radius relationship of \citet{ChenKipping2017}. The mass-radius relationship uses a piece-wise linear function which links mass and radius over scales from small rocky planets to stars. It is a rapid scheme that is included by default in the NASA Exoplanet Archive for the Planetary Systems with Composite Parameters (PS-CP) Table to provide an estimate of one of missing mass or radius.

The scatter in Figure~\ref{fig:fd_transit_mass} is generally higher than that for the complete dataset in Figure~\ref{fig:cd_transit_mass} and reflected in a higher average error, $\epsilon$. This is unsurprising, as many planets now have less observed properties from which to estimate the mass which is resulting in higher errors. The algorithm with the lowest overall error is the \kNNKDE, with the GAIN once again performing most poorly. With the inclusion of the full archive to now assist the imputation, the error for the original planets in the test set of TLG2020 (denoted by colored dots) has also changed. For \kNNKDE, MissForest and MICE, the addition of the full dataset has improved the mass imputation for the original test planets, but the error has actually increased significantly for $k$NN-Imputer and GAIN. In the case of $k$NN-Imputer, the inclusion of the full dataset is resulting in a relatively strong bias towards predicting an average mass for the high mass planets. This is visible as a horizontal plateau at around a Jupiter mass. The same plateau is also visible in the MICE algorithm and, to a lesser extent, in the \kNNKDE plot. The $k$NN-Imputer algorithm also shows evidence for a second plateau for lower mass planets, effectively selecting one of two average masses for the majority of the planet population, which explains the high average error. 

The emergence of the averaged plateaus depends on how each algorithm is imputing the single point missing values. Shown in the upper-left of Figure~\ref{fig:fd_transit_mass}, the \kNNKDE calculates imputed values by taking the mean of the probability distribution. This mean is affected by the number of neighbors the algorithm draws upon for the distribution. As described in section~\ref{sec:method}, the number of neighbors is determined by the newly added hyperparameter $N_{\rm cap}$. The value of $N_{\rm cap}$ needs to be large enough to accurately sample the parameter space in the vicinity of the planet whose properties are being imputed and provide an informative probability distribution. But for a single point imputation, $N_{\rm cap}$ should also be sufficiently low that all included neighbors have observed properties close to those of the imputed planet. If the distribution draws from too many neighbors, the mean becomes influenced by potential outliers near the distribution tails, leading to broad averages. Conversely, a lower neighbor number visibly reduces any plateau formation, but at the cost of a higher scatter and higher average error $\epsilon$: this is the famous bias-variance trade-off. Here, a value of $N_{\rm cap}=20$, was selected as a compromise between error performance on the average imputed value, and an informative distribution. Note that an average imputed value is computed primarily for the purpose of code comparison. However, as seen in section~\ref{sec:completedata}, a more thorough and interesting exploration of the exoplanet comes from studying the origin of the probability distributions for the imputed value. If one is not interested in a single imputed value, then plateau formation is not a problem since it originates from averaging over the distribution of choices for the mass.

Surprisingly, the $k$NN-Imputer (upper middle panel of Figure~\ref{fig:fd_transit_mass}) shows a much stronger plateau than the \kNNKDE even though the scheme uses less neighbors ($k=15$, see section~\ref{sec:method}). This is because the $k$NN-Imputer imputes the missing values for each planet individually, one after another, unlike the \kNNKDE that imputes all the missing values for each planet together. As selected neighbors are required to have observed values only for the missing property being currently imputed, imputing one property at a time allows the $k$NN-Imputer to select from a much larger pool of surrounding planets. This potentially unlocks access to more relevant neighbors, but also increases the risk for biased imputation, where the same neighbors are repeatedly used in denser areas of the parameter space. This is what is happening with the formation of the two mass plateaus at Jupiter-sized planets and at around 10\,\ME. Increasing the neighbor number for the $k$NN-Imputer would lead to even stronger plateaus. Visually removing the averaged plateaus requires dropping below $k=10$ neighbors for imputation, but the scatter is much higher with such a low number of neighbors, and the overall error also higher (lower bias implies higher variance in the prediction).

The apparent plateau for the imputed planet masses by the MICE algorithm (lower middle panel of Figure~\ref{fig:fd_transit_mass}) is also a consequence of a dense gas giant parameter space region. However, this problem cannot be addressed here as MICE has no hyperparameter to modify its behavior. Because MICE uses linear regressions to estimate missing values, the dense collection of observed masses in the Jupiter-sized regime draws mass estimates towards the broader average. In particular, MICE struggles to predict planet masses above 1,000\,\ME. That said, it is worth noting that the average plateau effect does not happen in the super-Earth regime, probably because the observed masses span a broader range, and also because other observed properties are more diverse in the super-Earth regime than with hot Jupiters.

Interestingly, MissForest do not show any plateau (upper right panel of Figure \ref{fig:fd_transit_mass}). As described in section~\ref{sec:method}, Random Forests--that are used by MissForest for regression--are considered the state-of-the-art for tabular data thanks to the precision of decision trees at the core of Random Forests methods. While single decision trees tend to overfit part of the parameter space, such as the Jupiter-sized planets, Random Forests leverage many decision trees and aggregates their predictions. This leads to much less bias (hence the absence of plateau) but potentially higher variability in predictions, especially in sparser areas of the parameter space. This can be see in the scatter for imputations at sizes in between the main super Earths and gas giant groups, and past 1,000\,\ME. 

Similar to section~\ref{sec:completedata}, GAIN produces a very poor imputation (lower left panel of Figure~\ref{fig:fd_transit_mass}). Because GAIN does not directly use the observed values in the dataset, but rather tries to mimic them, heterogeneous and potentially inconsistent values end up being proposed for imputation, which leads to a large scatter.

The mass estimates obtained via the M-R relationship of \citet{ChenKipping2017} as computed by the Planetary Systems with Composite Parameters (PS-CP) show the overall higher error (see lower right panel of Figure~\ref{fig:fd_transit_mass}). But this is mainly because the mass-radius piece-wise linear relationship is not invertible in the range 11.1\,\RJ to $14.3$\,\RJ, which despite being a small radius range, includes most Jovian worlds and spans over masses from 85 to 35,000\,\ME, obviously leading to poor results in that range. For planets with masses below 85\,\ME, the algorithm performs well and has one of the tightest estimates for the very small planets.

\begin{figure*}[!ht]
    \centering
    \includegraphics[width=0.99\textwidth]{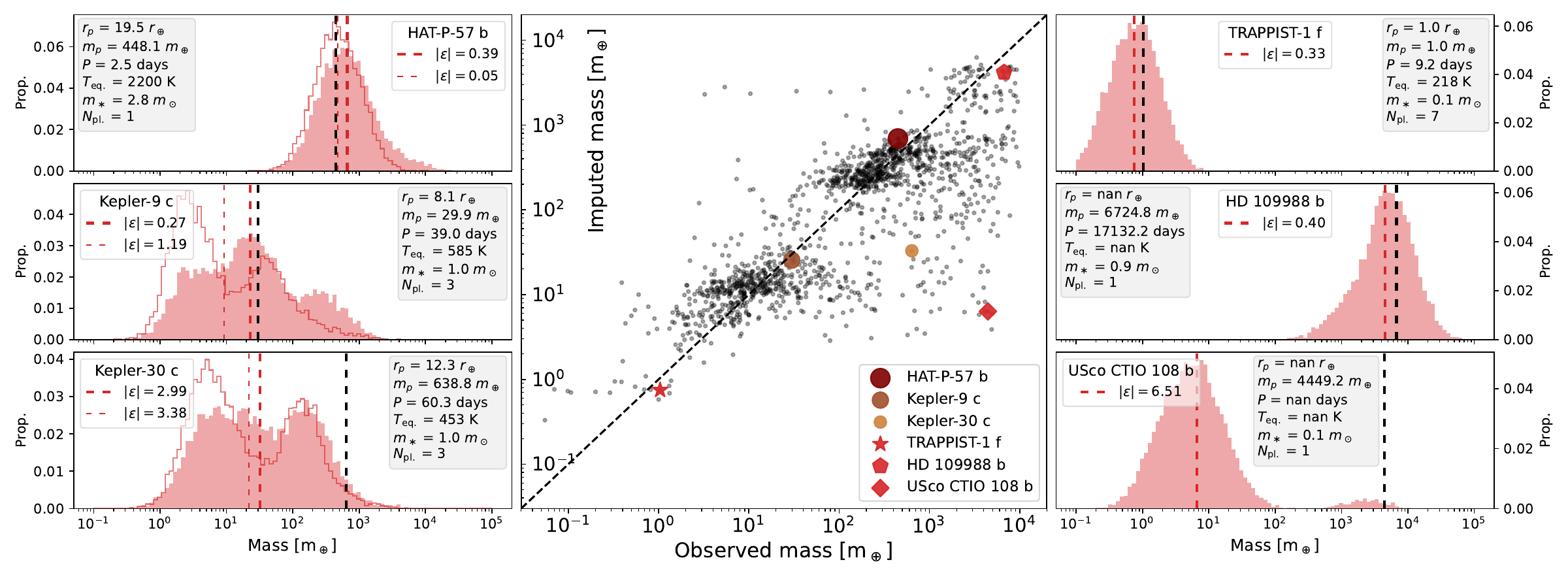}
    \caption{Six distributions for imputed mass values with the \kNNKDE in the ``transit regime" where planet mass is concealed and imputed. The three profiles on the left are the same planets as in Figure~\ref{fig:cd_transit_profiles}, whose mass was originally imputed using the complete properties dataset. The distribution using the complete properties dataset is shown outlined in red for comparison (note that in some cases, the planet properties have been updated since TLG2020. For the correct comparison in this case, the red outline shows the distribution using the complete properties dataset for these updated values). The error for Kepler-9c has substantially dropped with the use of the full archive dataset, but Kepler-30c is still an unusual planet. The three profiles on the right were selected as interesting examples: TRAPPIST-1 f is an Earth-sized planet, and one of the smallest in the dataset, HD 109988b is one of the most massive planets in the dataset and does not have a radius observation, while USco CTIO 108b has one of the highest errors. The central plot is a reproduction of the \kNNKDE case in the top left panel of Figure~\ref{fig:fd_transit_mass}, showing the location of the six planets.}
    \label{fig:fd_transit_profiles}
\end{figure*}

\subsubsection{Mass distributions in the transit regime: full archive dataset}
\label{sec:fd_transit_distrib}

Figure~\ref{fig:fd_transit_profiles} looks at the mass probability distributions for six planets from Figure~\ref{fig:fd_transit_mass} with observed but concealed mass values that have been imputed by the \kNNKDE algorithm. The left-hand three panels show the mass distributions for the same planets as shown in Figure~\ref{fig:cd_transit_profiles}, but now imputed making use of the full archive dataset rather than the smaller complete properties dataset. Note that newer observations have been conducted between the development of this paper and TLG2020, resulting in a few planets having updated measured properties in the newer dataset utilized in this section. None of these changes have strongly altered the distribution shape for the three planet comparisons, HAT-P-57b, Kepler-9c and Kepler-30c. However, in order to accurately assess the impact of drawing on the full archive dataset for imputing properties, the mass distribution for each planet with the updated measured values using the complete properties dataset of section~\ref{sec:completedata} has been recalculated. The updated distributions are shown as a red outline on the left-hand panels in Figure~\ref{fig:fd_transit_profiles}.

\paragraph{HAT-P-57b, Kepler-9c \& Kepler-30c: Dataset comparison} Interestingly, while there is an overall improvement in the error for the planets also present in the complete dataset, the error for HAT-P-57b has slightly increased. This is due to the addition of more massive planets present in the full dataset, a number of which orbit around higher mass stars and have higher equilibrium temperatures similar to HAT-P-57b. HAT-P-57 is a massive A-type star that is an unusual planet host in the current archive, and one where it would be easier to detect a high mass planet. The inclusion of these higher mass planets in the dataset therefore slightly extends the high mass tail in the imputed mass distribution for HAT-P-57b, giving a raised average mass value. However, the change is small with the observed mass value still lying at the peak of the imputed distribution.

By contrast, the mass prediction for Kepler-9c has substantially improved. While the previous distribution based on the smaller complete properties dataset had a bimodal shape that proposed two possibilities for the most likely planet mass, Figure~\ref{fig:fd_transit_profiles} shows evidence of three modes, although less distinctly separated than in Figure~\ref{fig:cd_transit_profiles}. The middle of the three new modes has the highest peak, and lies at the observed value for Kepler-9c. To the left of the central mode is a lower mass option that peaks around 3\,\ME. This mode lies in the same location as the stronger of the  bimodal peaks in the previous distribution. On the right of the distribution sits a third high mass mode at about 1\,\MJ that is at a slightly higher mass than the previous high mode peak but with much lower probability than the other two peaks.  

The addition of the third more accurate mode in the mass distribution of Kepler-9c is the inclusion in the full archive dataset of more planets with observed mass and radius between the two most common classes of gas giant and super Earth. This can be seen by the addition of light gray dots on the planet mass versus planet radius plot in Figure~\ref{fig:pairplot}. The broad profile still indicates that Kepler-9c is somewhat unusual, but the highest probability from the larger dataset does provide the correct solution. 

The less distinct peaks for Kepler-9c push away from the concept of very distinct planet classes and point towards a continuum of planet sizes created by a multitude of evolutionary pathways during planet formation. This is seen again in the new distribution for Kepler-30c. The sharply peaked bimodal mass distribution in Figure~\ref{fig:cd_transit_profiles} has softened to a broader spread of masses. The two original peaks are still visible at the same locations as in the previous distribution, but more options now sit in-between. However, in this case, the extra options have not improved the mass imputation which remains significantly lower than the observed mass. This is because the 60\,day orbit remains relatively rare, even in the full archive, and even fewer planets at that orbit have high masses, as can be seen in the pairplot of orbital period and planet mass in Figure~\ref{fig:pairplot}. 

\paragraph{(Low error) TRAPPIST-1f: Typical for a rocky world} The right-hand column of Figure~\ref{fig:fd_transit_profiles} shows three mass profiles for planets not studied in section~\ref{sec:completedata}. In the case of TRAPPIST-1f, the planet has a complete set of observed properties for the six considered parameters, but was in the training (not test) set for the mBM. The other two planets in that column have incomplete properties and so could not be included in the prior complete properties dataset. 

TRAPPIST-1f is one of seven approximately Earth-sized planets orbiting a low mass M-dwarf star \citep{Gillon2016}. The planet orbits in the so-called habitable zone, which would allow surface liquid water to persist if the planet hosted an Earth-like environment \citep{Kasting1993, Kopparapu2013}. Small worlds on temperate orbits are still relatively rare, as can be seen in Figure~\ref{fig:pairplot}, as are planets around very low mass stars and in systems with more than six known worlds. The \kNNKDE therefore finds a fairly loose collection of neighboring planets, many of which orbit the more commonly observed G-type stars. Despite this, the neighbors are in systems of six, seven or eight planets and are consistently low mass. Six of the nearest and most highly weighted neighbors are unsurprisingly the other planets in the TRAPPIST-1 system, which have very similar sizes due to the ``peas in a pod" effect discussed in section~\ref{sec:data}. The result is an accurate mass estimate, with the algorithm certainty demonstrated by a single, fairly narrow peak. 

As can be seen in the central panel of Figure~\ref{fig:fd_transit_profiles}, planets with observed masses significantly below that of TRAPPIST-1f have overestimated masses. This is likely due to an absence of any close neighbours, forcing the \kNNKDE to look exclusively to higher mass planets for guidance. Weighting close neighbors assists the \kNNKDE with handling planets in a sparse area of the parameter space, but it inevitably struggles if the planet is close to unique amongst current observations. As previously seen, uncertainty in the imputation is evident in the broadness of the resulting distribution. 

\paragraph{(Low error) HD 109988b \& (high error) USco CTIO 108b: Planets with sparse observations} The last two mass imputations in Figure~\ref{fig:fd_transit_profiles} are not technically a replication of a transit observations, as neither planet transits its host star and therefore have no observed radii measurements. Without an observed radius, the physical size of both HD 109988b and USco CTIO 108b therefore could not be estimated either by the previous mBM code in TLG2020, or by mass-radius relationships \citep[e.g.][]{Weiss2014, ChenKipping2017}. This therefore makes the planets an interesting test of the performance of the \kNNKDE algorithm.

The imputed mass for HD 109988b has a low error, with a distribution sharply peaked around the observed mass and a tail that extends more towards lower masses. With a mass over 21\,\MJ, HD 109988b is more properly a wide-orbit brown dwarf \citep{Feng2022}. A population of such celestial bodies with masses exceeding 1000\,\ME and orbital periods over 1000\,days are present in the full archive dataset, but entirely absent in the complete properties dataset due to the the difficulty in measuring the full set of six properties. From the pairplot in Figure~\ref{fig:pairplot}, HD 109988b can be seen to be typical of that distant, massive population of planets, explaining the successful imputation of its mass despite relatively few observed values. The low mass tail on the distribution is actually due to the planet's missing radius observation, which requires the \kNNKDE algorithm to search for neighbors with both measured planet radius and mass. Such coupling promotes consistency between all planet measurements, but reduces the potential neighbors in this planet group where most planets have been discovered by either the radial velocity or direct imaging methods, and do not transit. Relaxing that requirement might have tightened the profile in this case, but risked an inconsistent set of planet properties. 

The final panel in Figure~\ref{fig:fd_transit_profiles} shows the mass distribution for another very massive planet close to the deuterium burning limit, USco CTIO 108b \citep{Bejar2008}. This planet was discovered through direct imaging and unlike HD 109988b, does not have a measurement for the orbital period. This leaves only the stellar mass and number of planets from which to impute a mass estimate. The stellar mass is also exceptionally low, significantly less than the majority of known host stars. As a result of the high number of missing values, and sparse parameter area around the known stellar mass, the resulting imputation is challenging. The twenty closest neighbors identified by \kNNKDE that have the required four missing properties are mainly situated in the super Earth region of Figure~\ref{fig:pairplot}, due to smaller planets being slightly more commonly found around low mass stars. This creates the primary distribution peak at about 7\,\ME, with only two neighbors indicating that the higher (correct) mass is possible.  

The struggle with imputing properties for planets with a low number of measured values is not surprisingly, and evidence of this can also be seen in the central plot of Figure~\ref{fig:fd_transit_profiles}. A horizontal line of planets can be seen with imputed values all just below 10\,\MJ (3000\,\ME). These are planets all detected via gravitational microlensing, and similarly have only the mass of their host star and number of planets in the system from which to impute a mass value. 

Despite the stronger peak at 7\,\ME, a manual inspection of the mass profile for USco CTIO 108b would likely have resulted in the small high mass peak being consider the more likely value. This is because the algorithm does not use knowledge about the detection technique when estimating values. Such information was intentionally excluded to avoid the use of non-physical trends in the exoplanet demographics. However, in this case, the imaging detection points to a massive world. This is an additional example of where the distribution is more valuable than a single value imputation.

\subsubsection{Mass and radius prediction in the RV regime: full archive dataset}
\label{sec:fd_rv}

\begin{figure*}[!ht]
    \centering
    \includegraphics[width=0.95\textwidth]{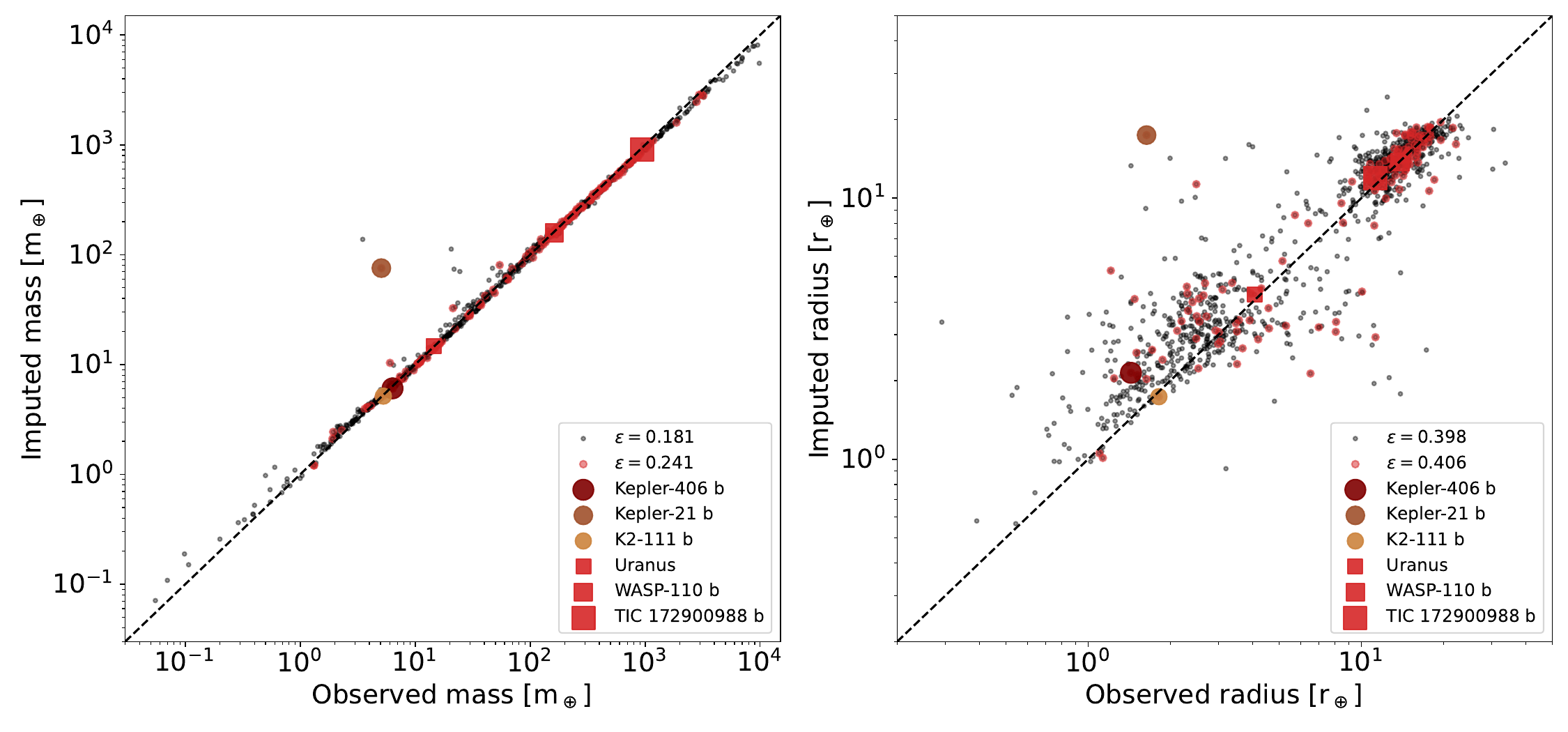}
    \caption{Test results for the \kNNKDE algorithm when using the full archive dataset and treating each planet as a radial velocity observation, with concealed and imputed planet radius and mass values, weighted by a given minimum mass measurement. The plots show the results for the mass imputation (left) and radius imputation (right), after the distribution has been weighted by the minimum mass. Black dots show all planets in the full archive dataset, while red dots show the planets that were also in the test set for the complete properties dataset. The profiles for the planets identified in the legend are shown in Figure~\ref{fig:fd_rv_profiles}.}
    \label{fig:fd_rv_mass_radius}
\end{figure*}

Figure~\ref{fig:fd_rv_mass_radius} now looks at the performance of the \kNNKDE algorithm in imputing both planet mass and radius based on the remaining four properties in the dataset, together with a minimum mass value, leveraging the full archive dataset. This is equivalent to treating each planet that has a measured mass and radius as if it were a radial velocity detection, and assessing code performance by imputing those (concealed) properties. As in section~\ref{fig:cd_rv_mass_radius}, this is a test that can only be performed with the \kNNKDE algorithm, since a distribution of imputed mass values is needed to convolve with a minimum mass measurement, as described in section~\ref{sec:method_minmass}. The black dots in Figure~\ref{fig:fd_rv_mass_radius} show the results for imputing the mass and radius for the 1,081 planet in the full archive dataset with both measured values, while the red dots indicate planets that were also in the complete properties dataset of TLG2020.

As for the transit regime comparison in Figures~\ref{fig:cd_transit_mass} and \ref{fig:fd_transit_mass}, the average error for the planets common to both datasets has decreased with the use of the full archive dataset for the imputation. However, there is now an even lower error when averaged over all the planets in the full dataset, whereas in the transit regime test, the larger dataset had a higher average error. This error reduction in the radial velocity regime test reflects that the planets in Figure~\ref{fig:fd_rv_mass_radius} have less variability in the data available for imputation, with all entries having a minimum mass measurement but no radius or true mass available. In the equivalent transit regime test in Figure~\ref{fig:fd_transit_mass}, a planet radius measurement was available for part of the archive, resulting in a mix of imputations for planets with and without a size guide. Since the remaining four parameters (orbital period, equilibrium temperature, stellar mass and number of known planets in the system) are more commonly measured than planet radius, this variation in planet radius measurements for the transit regime results in more scatter. As with the radial velocity regime test for the complete properties dataset in Figure~\ref{fig:cd_rv_mass_radius}, the relation between the imputed mass and observed mass is very tight, due to the value of a minimum mass in guiding the true mass value. Notably, there is no averaged plateau for either the mass or radius imputed values. This is mostly due to the observed minimum mass value, whose convolution with the mass distribution prevents broad averaging over the full imputed distribution.
  
\begin{figure*}[!ht]
    \centering
    \includegraphics[width=0.95\textwidth]{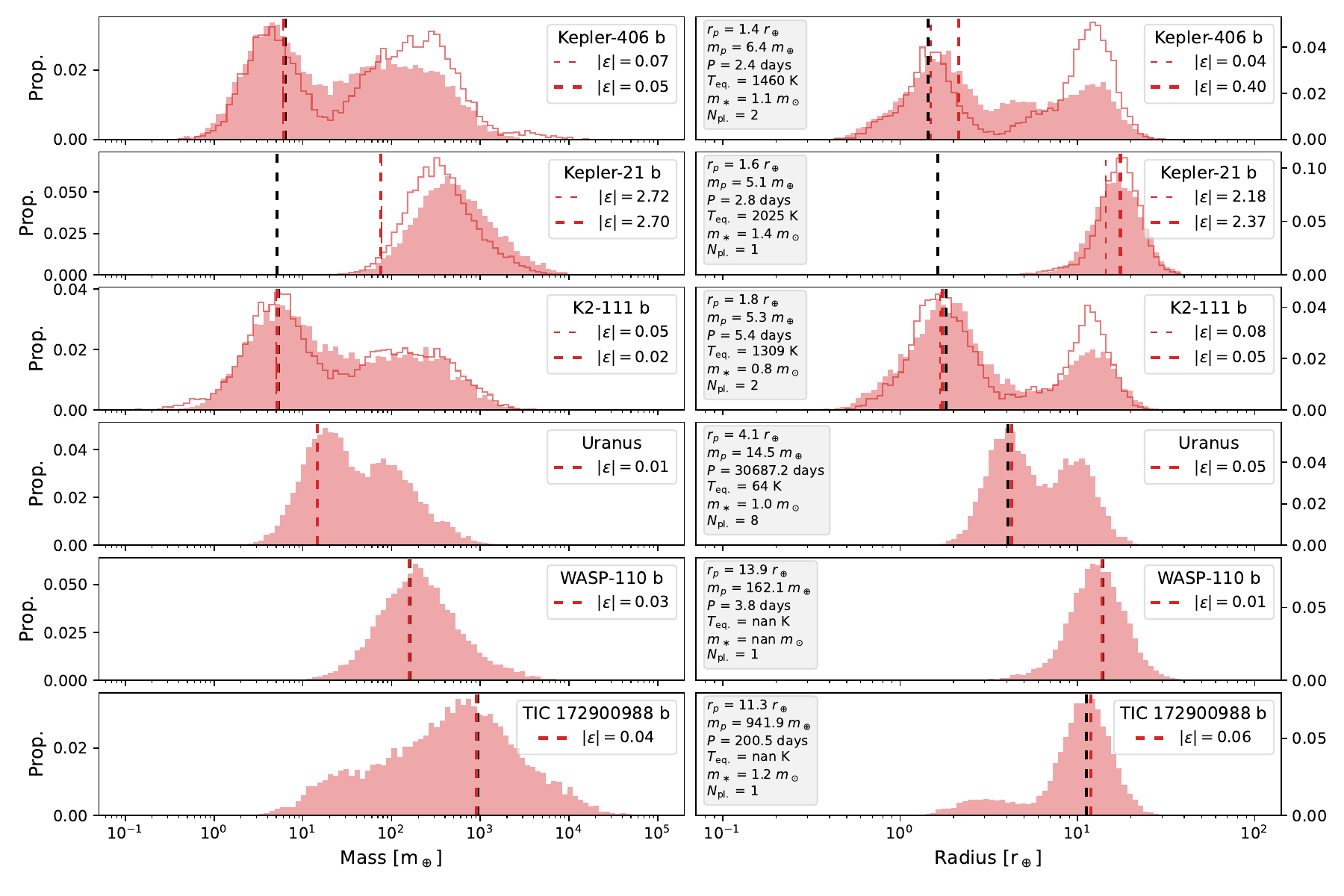}
    \caption{Distributions for the imputed mass and radius values calculated with \kNNKDE for the planets highlighted in Figure~\ref{fig:fd_rv_mass_radius} using the full archive dataset. The top three planets, Kepler-406b, Kepler-21b and K2-111b, are the same planets whose distributions were shown in Figure~\ref{fig:cd_rv_profiles} when imputed using the smaller complete properties dataset. The bottom three planets show interesting cases where there was a successful mass imputation. Distantly orbiting Uranus is an outlier in the planet archive, but the presence of Neptune acts as a strong mass guide. WASP-110b is recognisably a hot Jupiter despite no stellar or temperature information. The observed mass of TIC 172900988b is uncertain, and the planet does not belong to an obvious class. This is indicated by the broad distribution, although the peak agrees with the observed estimate.
    The red outline shows the profile when calculated using the complete properties dataset, while the histogram is the profile with the full archive dataset. (As previously, the imputation with the complete properties dataset has been updated with the latest observations for the planet.) The average imputed value is shown by a red dashed line (thin dashed line for the complete properties dataset), with the error in the legend. The black dashed line is the observed mass and radius. The last three profiles are for planets that are not in the test set for the complete properties dataset.
    }
    \label{fig:fd_rv_profiles}
\end{figure*}

\paragraph{Kepler-406b, Kepler-21b \& K2-111b: Dataset comparison} The distributions for the imputed masses and radii for particular cases highlighted in Figure~\ref{fig:fd_rv_mass_radius} are shown in Figure~\ref{fig:fd_rv_profiles}. The top three profiles show the same planets as in Figure~\ref{fig:cd_rv_profiles}, but now with their properties imputed using the full archive rather than the smaller complete properties dataset. As with Figure~\ref{fig:fd_transit_profiles}, there have been new observations of the planets since TLG2020 dataset was made. The red outline therefore shows the distribution when the complete properties dataset is used to impute the mass and radius based on any updated properties. The average mass and radius values (the imputed values in Figure~\ref{fig:fd_rv_profiles}) are shown as red dashed lines, with the epsilon error in the legend. The thin dashed line is the result when using the complete properties dataset.

In contrast to the change in the profile shape seen in Figure~\ref{fig:fd_transit_profiles} for the transit regime test, the use of the full archive dataset has a smaller effect on the profiles for Kepler-406b, Kepler-21b, and K2-111b. This is not very surprising. The \kNNKDE code performs the imputation of missing values by searching for neighbors in the planet parameter space that have measured values for all missing properties. In the case of the transit regime test, this often involves only requiring neighbors to have a measured mass value. The choice for potential neighbors therefore significantly expands when the full archive is utilized. However, in the radial velocity regime test, neighbors must always have both a  mass and radius measurement. The neighbors involved in the imputation are therefore often also in the complete properties dataset (since planetary mass and radius have the highest missing rate as seen in Table~\ref{table:missingrates}), so there is a stronger overlap between the neighbor selection for the two datasets. This constraint is necessary to ensure that the imputed values are consistent values across properties, but does mean that it is harder to fully leverage a large incomplete database when imputing multiple values, as in the radial velocity regime. The $k$NN-Imputer algorithm does not have this constraint, but generally performs more poorly than the \kNNKDE, as seen in Figures~\ref{fig:cd_transit_mass} and \ref{fig:fd_transit_mass}. This issue will be returned to in section~\ref{sec:extended}. 

Despite the above restriction, the expansion to the full archive dataset has adjusted the distribution for Kepler-406b with a decrease in the probability of the higher mass and radius peak in the bimodal profile, and now more strongly favor a correct, smaller sized planet. When combined with the minimum mass distribution that also points to the smaller mass peak, this reduces the error for the average imputed mass value. The peak value of the radius distribution also agrees with the observed value, although the average value gives a higher error. The shift in the average is because the use of the full archive has flattened the higher mass peak and increased the probability of planets at intermediate sizes between the two peaks. While the probability of these planet sizes remains reasonably low, it is sufficient to push the average towards higher options. This change in profile shape is due to the wider range of radii values now observed for planets on short orbital periods with high effective temperatures, where heating can cause the atmosphere of planets to inflate \citep{Enoch2012}. In the case of Kepler-406b, the planet's average density is 11.8\,gcm$^{-3}$, indicating a rocky planet without a thick atmosphere that would have the potential to cause variations in radius \citep{Marcy2014}. Consistent with this, the observed mass and radius for Kepler-406b lie near the main peak of the distribution at the low mass and small size end.

Conversely, the distributions for Kepler-21b show almost no difference between the imputations using the two datasets. The planet has a high error, as the observed value is substantially lower than that predicted by the code. In this case, the planet's very high equilibrium temperature and massive host star means that it still remains an outlier in the full planetary demographics. This would be indicated by a minimum mass measurement, which would require an orbital inclination of less than 1\,degree to be consistent with the peak value.

The distribution for K2-111b is significantly changed from the profile discussed in Figure~\ref{fig:cd_rv_profiles} due to the discovery of a second planet orbiting K2-111 since the original complete properties dataset was created. As discussed in section~\ref{sec:completedata}, this significantly increases the chances of a planet being smaller than a gas giant, and the distribution shape for K2-111b is now bimodal when either the complete properties or full archive dataset is used in the imputation, as seen by the red outline in Figure~\ref{fig:fd_rv_profiles}. While the mass profile shows only a small change when the full archive is employed, the correct smaller radius value is more strongly favored with the full archive. This is from the increase in smaller planets at high equilibrium temperature. The epsilon error for this planet is now extremely low, with the use of the full archive dataset offering the best match.

The next three profiles show planets that were not in the test set for the complete properties dataset. Our own Solar System's Uranus has a complete set of observed properties, but was in the training set for the mBM algorithm. Both WASP-110b nor TIC 172900988 b were only discovered recently in 2021, and neither has complete properties that would allow them to be included into the complete properties dataset. 

\paragraph{(Low error) Uranus: An outlier with a close neighbour} As an outer planet in our own Solar System, Uranus is an outlier in its orbital period and equilibrium temperature. The majority of exoplanets found with periods longer than 10,000 days are usually young massive planets discovered by direct imaging. However, despite being quite an extreme outlier, the \kNNKDE algorithm accurately predicts the planet mass, with the observed value sitting close to the peak of the distribution, which is supported by the minimum mass, with the combined distributions placing the imputed value at the \kNNKDE peak. This is due to the presence of Neptune, which sits very close to Uranus across the parameter space and therefore becomes the most strongly weighted of the twenty neighbors used in the \kNNKDE imputation. The remaining neighbors span a range of masses, but are more weakly weighted. The gas giant population can be seen as a second, lower peak at high mass and radius in the distribution (driven mainly by Saturn and Jupiter), and smaller planets are selected due to our Solar System having high planet multiplicity (although very weakly weighted), which usually indicates smaller worlds. Similar to TRAPPIST-1f, the accuracy of the imputed value for Uranus demonstrates the importance of using a weighted neighbor scheme when performing probability density estimation; planets in sparser areas of the parameter space can retain accurately estimated properties by favoring the small number of similar discoveries, rather than a non-weighted average which risks covering a large range of the parameter space. For planets in very sparse areas of the parameter space, it is worth noting that the imputation may depend on just two or three close neighbors. 

\paragraph{(Low error) WASP-110b: A recognisable planet class} WASP-110b belongs to the hot Jupiter population, with a mass between that of Jupiter and Saturn and an inflated radius larger than Jupiter \citep{Nikolov2021}. There is no stellar mass nor equilibrium temperature recorded in the NASA Exoplanet Archive, so the imputed mass and radius values are based on the planet's orbital period and number of known planets in the system. However, the hot Jupiter population is densely clustered in single planet systems at orbits of a few days, allowing the \kNNKDE algorithm to make an accurately imputed mass and radius with a  profile clearly focused on the gas giant population that is only slightly broader than that for hot Jupiter HAT-P-57b in Figure~\ref{fig:fd_transit_profiles}, where more data is available. The convolution with the minimum mass does not move the imputed value from the peak of the distribution. 

\paragraph{(Low error) TIC 172900988 b: An difficult observation} TIC 172900988 b is a more complex case. The multi-Jupiter mass planet orbits in a binary star system with a circumbinary orbit that circles two stars of similar mass \citep{Kostov2021}. The actual mass of TIC 172900988 b is uncertain, with estimated values extending from $824$\,\ME$\le M_p \le 981$\,\ME, due to multiple solutions for the orbital properties. An analysis with an algorithm like the \kNNKDE presented here is a possible path to reducing the uncertainty, by an independent estimate of the planet properties based on similar planets in the multi-dimensional parameter space. This made the planet an interesting case study. However, its properties do also present challenges for the machine learning imputation. As there are very few binary systems in the NASA Exoplanet Archive, the stellar host number was not included as a property in the database that can be utilized by the algorithms. Moreover, the complexity of a temporally varying equilibrium temperature for the planet means that this parameter is additionally omitted. The stellar mass in this case was that of the primary star. Even with these challenges, the overall imputed values for the planet's mass and radius as seen in Figure~\ref{fig:fd_rv_profiles} are a close match to the recorded observed value of 942\,\ME and 11.3\,\RE, with the observed value lying close to the peak of the distribution, even without the minimum mass guide. The orbital period of TIC 172900988 b is long for the majority of planets in one planet systems, which is dominated by the close-in hot Jupiter population. The selected neighbors are therefore not as tightly gathered compared to WASP-110b, and have a broader range of values, with no very close neighbors dominating the weighting (differing from planets such as Uranus). This is reflected in the wide distribution width for the mass. The resultant distribution suggests that the recorded value is the most likely one for the planet.

\subsubsection{Summary: full archive dataset} 
\label{sec:fd_summary}

Overall, the extension to the full archive dataset improves the mass imputation when comparing planets also in the complete properties dataset. In particular, many planets that are unusual cases in the complete properties dataset are not outliers when all planets are included. This emphasises the value of including incomplete data from all observation techniques. In the next section, we consider the effect of adding more information on the planet properties to the dataset.

\subsection{The extended dataset: eight properties}
\label{sec:extended}

The ability to utilize an incomplete database opens up the possibility to leverage information from more planet properties for the imputation of missing values. Previously, the database had been restricted to six properties that were selected to be informative about the nature of the planet, while also having sufficient observed values from which to build a training set of complete properties for the mBM network. The algorithms developed for this paper removes that restriction, and allows additional potentially informative properties to be included in the imputation of missing values, even where there is a high fraction of missing values in the archive.

In this section, the full archive database is extended to include the host star metallicity and the planet orbital eccentricity. As discussed in section~\ref{sec:data}, trends between pairs of variables indicate that both properties may assist the imputation of missing values. Note that as all properties are equally weighted in algorithms such as the \kNNKDE, adding uninformative properties to the imputation would risk lowering the accuracy of the imputation results. Both stellar metallicity and orbital eccentricity were missing from the previous datasets due to their low completeness in the NASA Exoplanet Archive (see Table~\ref{table:missingrates}). However, although non-complete datasets can be leveraged by the algorithms presented in this paper, a very low number of observed values does still present challenges. In particular, the \kNNKDE algorithm requires twenty neighbors in the parameter space to a planet whose properties are to be imputed, all of which must have observed values for all missing properties. If one or more property is rarely observed, the distance to the twenty neighbours can become very large, and the imputation proportionately poorer. This issue was mentioned in section~\ref{sec:fd_rv}, but now becomes a more serious problem for the algorithm due to the low completeness of the extended dataset. To tackle this issue, the \kNNKDE algorithm was adjusted so the user could choose whether to impute the value of all missing properties, or just a subset. For the properties which would not be imputed, the algorithm dropped the requirement that the selected neighbor planets had to have this property measured. This allows the low-completeness stellar metallicity and orbital eccentricity to be used to define which neighbors are closest within the eight dimensional parameter space, but accept neighbors for the imputation with only requested parameters for imputation. For consistency with the previous two sections, the \kNNKDE algorithm imputed all of the original six planet properties where missing, but did not impute the stellar metallicity or orbital eccentricity, using these exclusively to select neighbors with relevant properties. 

The other four algorithms have also been dropped at this stage in the analysis. The \kNNKDE algorithm has been a top performer for the previous two datasets, and the information from the probability distributions that the \kNNKDE can create greatly exceeds that available from a single imputed value. 

\subsubsection{Mass prediction in the transit regime: extended dataset}
\label{sec:ed_transit}

\begin{figure*}[!ht]
    \centering
    \includegraphics[width=0.95\textwidth]{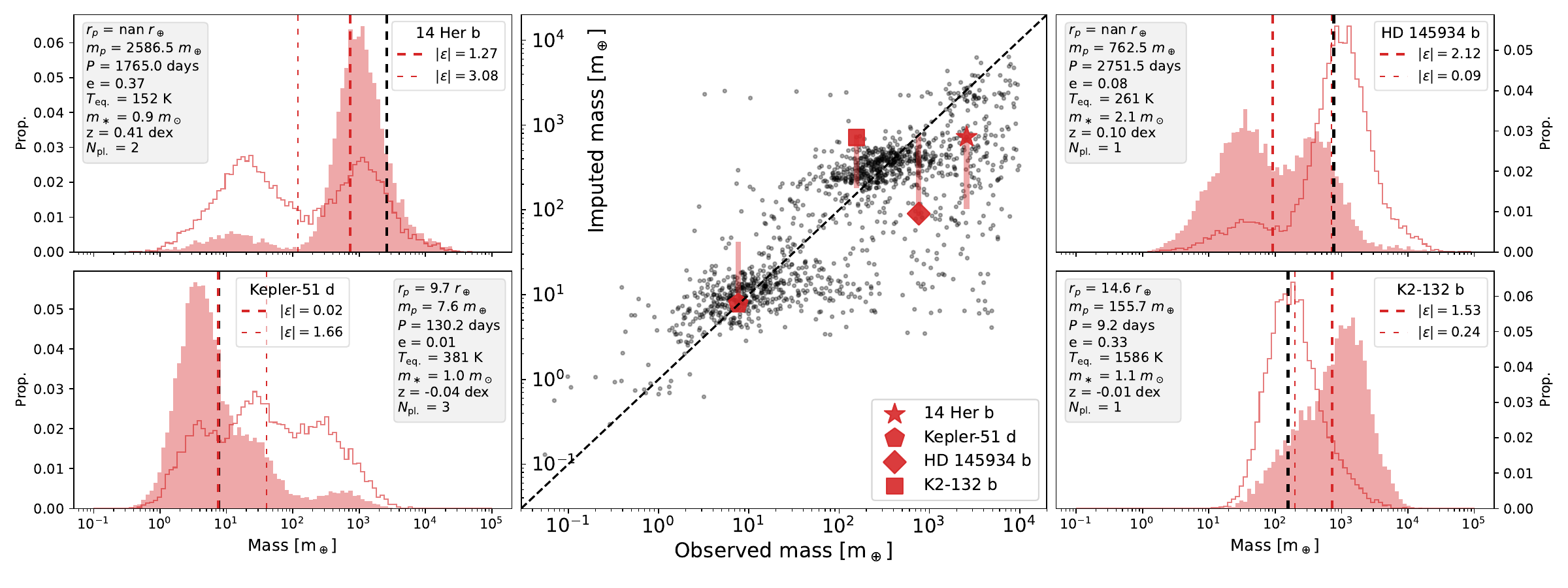}
    \caption{Test results when using the extended archive which leverages eight planet properties during the imputation. This is the transit regime test, where known planet mass values are concealed and imputed by the \kNNKDE algorithm, similar to what would be needed to impute mass for a transit observation. The $\epsilon$ error for the 1,426 planets plotted is $\epsilon = 1.502$, and when computed only over the planets that were in the test dataset of TLG2020, the error is $\epsilon = 0.840$. This demonstrates a small overall improvement in mass imputation accuracy. Distributions for four planets marked on the central plot are shown to the left and right, selected for the big change in their errors after the inclusion of the two extra parameters. The two planet profiles on the left show planets where the error decreased with the inclusion of two extra planet properties in the imputation. On the right, are two planets where the error degraded with the extra information. The red outline on each histogram is the distribution for the planet when using the previous six property dataset. The filled histogram is the distribution when using the eight property dataset. The location of these planets on the central mass imputation plot are shown with a red line indicating their original imputed value when using the previous six-property full archive dataset.}
    \label{fig:ed_transit_mass}
\end{figure*}

Figure~\ref{fig:ed_transit_mass} shows the results from the \kNNKDE algorithm for the transit regime test, where known mass values are concealed and estimated as might be required for imputing missing mass values in transit observations. This is the same test that was performed for the full archive dataset in Figure~\ref{fig:fd_transit_mass}, but now with eight properties including the planet orbital eccentricity and stellar metallicity used in the imputation. Note that the orbital eccentricity is included in the imputation where present, even if this would not normally be measured as part of a transit observation. Likewise, not all the planets plotted above have radius measurements. The epsilon error over the full dataset has decreased slightly with the addition of two extra parameters, moving from $\epsilon = 1.510$ (six parameter dataset) to $\epsilon = 1.502$ (eight parameter dataset), demonstrating a small overall improvement when adding additional information. A similarly small improvement can be seen on the subset of planets in the test set of TLG2020, going from $\epsilon=0.846$ (six parameter dataset) to $\epsilon=0.840$ (eight parameter dataset).

To take a closer look at how the extra information in the extended dataset can impact the planet property imputation, four planets are highlighted in the central plot in Figure~\ref{fig:ed_transit_mass} that have measured values for orbital eccentricity and stellar metallicity, and whose error values have changed significantly when using these two additional properties. The imputed mass value using the eight properties extended dataset is marked with the red shape, and a red trail indicates the location of the previous imputed value when utilising the six properties dataset. The profiles for each planet are shown flanking the central plot. 14 Her b and Kepler-51d shown on the left side of Figure~\ref{fig:ed_transit_mass} have greatly improved imputed mass values, whereas the imputed mass values for HD 145934 b and K2-132b deteriorated with the additional information.

\paragraph{(Low error) 14 Her b: A high eccentricity giant} 14 Her b (also known as HD 145675 b) is a massive gas giant on a Jupiter-like orbit with a period of 4.8 years around a Sun-like star \citep{Butler2003, Feng2022}. Discovered via radial velocity, this planet does not have a radius measurement, so its mass during this transit regime test is based on orbital period, equilibrium temperature, stellar mass and number of planets when using the six properties full archive dataset, and these four properties plus orbital eccentricity and stellar metallicity when using the new extended properties dataset. In the pairplot shown in Figure~\ref{fig:pairplot}, 14 Her b belongs to the cluster of high mass, long period planets that can be seen towards the top right in the planet mass versus orbital period plot. Notably, this group of planets rarely transit, so very few have radius measurements. This means that the \kNNKDE algorithm has to go outside this cluster for neighbors with measured values for the six main properties, stepping into the parameter space for the longer period gas giants down to super Earths. The result is a bimodal distribution when utilising the six properties full archive dataset, with peaks close to the measured gas giant value and at the super Earth mass of about 11\,\ME. With the inclusion of the orbital eccentricity and stellar metallicity, this degeneracy breaks, and the higher gas giant mass is strongly favoured by the \kNNKDE algorithm. This is because the relatively high orbital eccentricity at $e = 0.37$ is far more commonly found for high mass planets; a trend that was noticed in the discussion of the Pearson correlation coefficients in Figure~\ref{fig:pearson_corr_coeff}. Removing or lowering the weighting on planets with lower eccentricity measurements for the close neighbors therefore increases the algorithm's uncertainty that this is a gas giant.

\paragraph{(Low error) Kepler-51d: Around a low metallicity star} Kepler-51d is a second case where the addition of the orbital eccentricity and stellar metallicity has greatly increased the certainty of the \kNNKDE algorithm to favour a particular planet mass regime. In this case, the profile has changed from a fairly continuous distribution between a gas giant and rocky planet, to a strongly peaked profile at a few Earth masses. Kepler-51d is an unusual planet in the archive as it has a very low density \citep{Masuda2014}. Based on a radius measurement alone, a gas giant would be suspected. With the six properties available from the full archive dataset imputation, the multi-planet system and orbital period suggest lower masses may be equally probable. However, both the orbital eccentricity and stellar metallicity offer clues to narrow down the distribution. The orbital eccentricity for Kepler-51d is very low, which is common for a wide range of planet masses. However, many high mass planets with low eccentricity will probably be in tidal lock, on much shorter orbital periods than Kepler-51d. Moreover, Kepler-51 has a sub-solar metallicity, which slightly favors lower mass planets. This adjusts and re-weights the neighboring planets so that the lower mass becomes the strongly dominant peak. 

\paragraph{(High error) HD 145934 b: A misleadingly low eccentricity} On the right-hand side of Figure~\ref{fig:ed_transit_mass}, two profiles are shown for planets whose mass imputation significantly deteriorated with the additional information from orbital eccentricity and stellar metallicity. The profile for HD 145934 b moved from a strong (and correct) estimation that this was a gas giant, to an equal chance of both a gas giant and super Earth. Like 14 Her b, HD 145934 b belongs to the group of long period, high mass planets discovered via radial velocity that do not have radius measurements \citep{Feng2015}. However, the high mass of the star HD 145934 and the lack of a second planet in the system means that the six properties full archive dataset is confident that this is a gas giant. This certainty is likely upset by the low orbital eccentricity of HD 145934 b when the imputation is based on the eight properties extended dataset. Low eccentricity is expected for single planet systems with less scattering and also for planets with lower masses, as is seen in the data (see Figure~\ref{fig:pearson_corr_coeff}). The low eccentricity value therefore increases the parameter space distance from high mass neighbors that also have high eccentricity, and creates a more even probability between the two mass options. That said, a visual examination of the two peaks would likely result in the higher peak being correctly selected for the mass imputation rather than the average, due to gas giants being easier to detect on long periods.

\paragraph{(High error) K2-132b: An evolving system} The last of the four distributions in Figure~\ref{fig:ed_transit_mass} is K2-132b. The addition of orbital eccentricity and stellar metallicity has pushed the planet distribution to higher masses, taking the peak further from the measured value. K2-132b is an inflated gas giant that sits in a dense region of the six properties full archive parameter space. This produces a single, steep peak very close to the measured mass. However, the planet has a high eccentricity which is  more unusual for a large planet on a short orbit. The star evolution has been proposed as a reason for this high eccentricity, with tides on evolved stars causing a transient high eccentricity orbit for the host planet \citep{Grunblatt2018}. The inclusion of the eccentricity therefore selects more highly eccentric neighbors, which favors more massive planets. The result is a quite broad profile (indicating some uncertainty in the imputation) that is skewed towards higher masses. This might be avoided by including spectral type in the imputation, but the value is only recorded in the archive for a relatively small number of entries. K2-132 b is therefore actually an example of an outlier planet, which only appeared to be typical when considering a reduced number of properties.

\subsubsection{Mass and radius prediction in the RV regime: extended dataset}
\label{sec:ed_rv}

\begin{figure*}[!ht]
    \centering
    \includegraphics[width=0.95\textwidth]{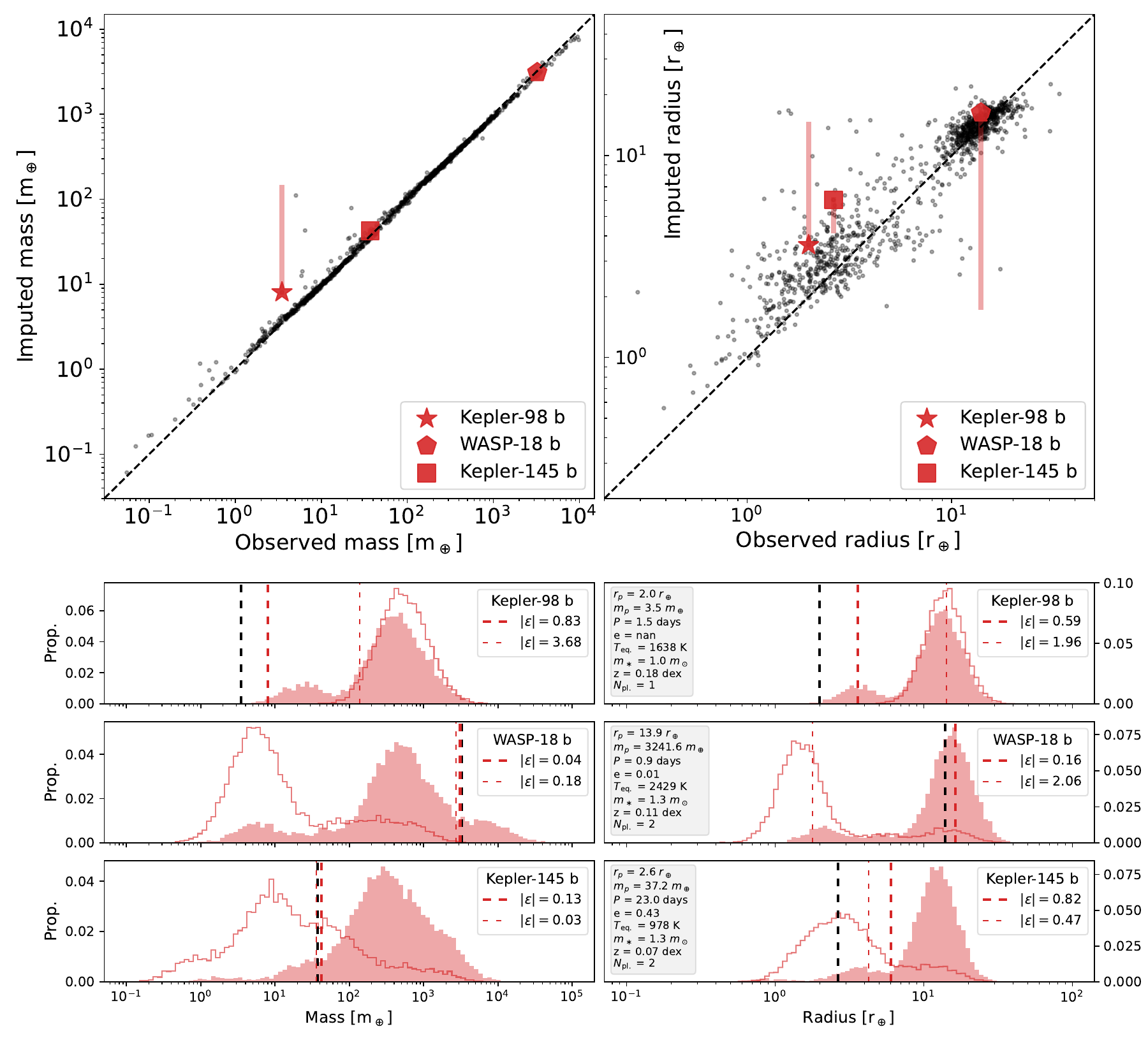}
    \caption{Test results when using the extended archive with eight planet properties for imputing both planet mass and radius with the inclusion of a minimum mass value, ``radial velocity'' with measured planet mass and radius values. The average error for the mass imputation is $\varepsilon = 0.157$, and the radius imputation has an average error of $\varepsilon=0.363$, showing a small improvement in accuracy compared with the previous six properties full archive data set. The imputation is explored further in three mass and radius distributions shown in the lower half of the figure, selected because their error value changed significantly with the inclusion of extra properties. The filled histogram shows the property distribution using the eight parameter extended dataset, while the red outline shows the distribution when imputed with the previous six properties full archive dataset. The average imputed value is shown by a thick red dashed line for the eight properties extended dataset and thin dash line for the six properties dataset. The error for both is shown in the legend. The black dashed line is the measured value. The planet measured properties are shown in the gray box. The location of these planets is also marked on the upper panels, with a red trail indicating the previous imputed result with the six properties full archive dataset.}
    \label{fig:ed_rv_mass_radius}
\end{figure*}

As with the previous two datasets, the imputation to estimate both the planet mass and planet radius when utilizing a measured minimum mass value can be evaluated, as would be common with a radial velocity observation. The result of imputing known masses and radii values in the exoplanet archive with the eight parameter extended dataset is shown in Figure~\ref{fig:ed_rv_mass_radius}. The error for the planet mass averaged over the 1,081 planets plotted after convolution with the minimum mass has lowered with the inclusion of the additional two planet properties from $\varepsilon = 0.181$ (six parameter dataset) to $\varepsilon = 0.157$ (eight parameter dataset), with the average error on the imputed radius also dropping from $\varepsilon=0.398$ (six parameters) to $\varepsilon=0.363$ (eight parameters). The imputation slightly worsens when considering the planets in the original TLG2020 test set, changing from $\varepsilon=0.241$ (mass) and $\varepsilon=0.406$ (radius) (six parameters) to $\varepsilon=0.275$ (mass) and $\varepsilon=0.396$ (radius) (eight parameters).

\paragraph{(Low error) Kepler-98b: A smaller possibility} Below the plots of the average imputed values, mass and radius distributions for three planets are shown that demonstrate a change in the profile shape due to the addition of the orbital eccentricity and stellar metallicity properties. Kepler-98b is a super Earth whose high error is substantially reduced by the inclusion of the additional two properties in the imputation \citep{Marcy2014}. Using the six property full archive dataset, the \kNNKDE algorithm believes the planet to be a Jupiter-sized gas giant based on the planet's orbital period, equilibrium temperature, host star mass and number of planets in the system. This is principally driven by the 1.5 day orbital period and single planet status, which is very common in the hot Jupiter population. The nearest neighbors to Kepler-98b therefore end up being universally gas giants. The orbital eccentricity is not recorded for Kepler-98b, so the eight parameter database leverages only the stellar metallicity. While higher metallicity stars are more likely to host gas giants that stars of lower metallicity, smaller planets are also commonly found in orbit. This appears to be enough to slightly expand the knot of nearest neighbours away from the exclusively hot Jupiter population and produce a second small peak at lower planet mass in the distribution. In combination with the minimum mass measurement that indicates a lower mass planet, the final mass estimate is closer to the measured value. In this case therefore, the added planet property has helped to diversify the nearest neighbors. 

\paragraph{(Low error) WASP-18b: A disrupted planet} Unlike Kepler-98b, WASP-18b actually is a hot Jupiter and it seems initially surprising that the planet would have a high error on its mass and radius with any chosen dataset. However, WASP-18b is a particularly massive planet at 10\,\MJ, on an orbit of less than 1 day. The planet is so close to its host star that tidal interactions are likely on the brink of destroying the planet \citep{Hellier2009}. With no mass or radius measurement to initially guide the \kNNKDE imputation, the ultra-short orbit and high equilibrium temperature, coupled with a second discovered planet in the system, suggests a rocky or super Earth size for the planet. Adding in the orbital eccentricity and stellar metallicity in the extended dataset significantly improves this imputation, reflecting the peaks of the distribution so that a high mass planet is favored. It is not immediately obvious why this would occur, as the tidal circulation of orbits means that low eccentricities is expected for both high and low mass planets with short periods. However, the exceedingly small eccentricity for WASP-18b, and the higher metallicity of the star, has increased the proximity of a group of low eccentricity, higher mass planets in the parameter space that can be seen in the top left corner of the planet radius versus orbital eccentricity pairplot in Figure~\ref{fig:pairplot}, causing these gas giants to be favored as close neighbors. 

As all properties are weighted equally in the \kNNKDE algorithm, the use of eight properties also lowers the significance of each individual property. This may have been advantageous in the case of Kepler-98b and WASP-18b, as it lowers the significance of the number of planets in the system which was misleading in these cases. 

\paragraph{(High error) Kepler-145b: A possibly misleading eccentric} Unlike the previous two planets, Kepler-145b is an example where the imputed planet mass and radius has degraded as a result of the extra properties in the extended dataset. Kepler-145b is a planet between Neptune and Saturn in size, around a massive F-type star on a 23 day orbit. When using the six properties full archive dataset, the \kNNKDE algorithm favors a planet of the correct radius but lower mass. This is not surprising when looking at the planet's location on the planet mass versus radius pairplot in Figure~\ref{fig:pairplot}, as the planet has a high density. Discovered by transit timing variations, there is a large error in the observed mass estimate, so it is possible that the recorded observed mass is an overestimate \citep{Xie2014}. When the dataset is expanded to eight properties, the imputed distribution peaks at a higher mass and radius than observed. This is driven by the high orbital eccentricity of the planet, which indicates a high mass world. It is maybe worth noting that the measured eccentricity is also in doubt, as \citet{VanEylen2015} who reported the measurement noted that the transits of Kepler-145b were too shallow for any meaningful constraints on eccentricity. The outer planet, Kepler-145c, is thought to be in a close to circular orbit, so it could be that the eccentricity for Kepler-145b is artificially high in this instance.

\subsubsection{Summary: extended dataset}
\label{sec:ed_summary}

The addition of the two extra parameters improves the average mass imputation for the planets in the full archive dataset, demonstrating that extra information has been valuable in estimating the planet mass. This is particularly true if the planet has a high or low value for the new property, which helps narrowing down the planets with overall similar properties to be used for the mass imputation. However, as the \kNNKDE weights all properties of equal importance, extra information that does not relate to mass risks diluting the impact of more relevant properties.

\section{Exploring planet classification}
\label{sec:generative}

In addition to imputing missing values in a dataset, the \kNNKDE algorithm can be used as a fully generative model to create an arbitrary large number of planets with completely synthetic properties that still maintain the statistics of the original dataset. This synthetic planet population will have a complete set of properties and can therefore be used to explore the planet demographics using algorithms to visualize clusters within the high-dimensional parameter space. While such statistically defined clusters are not guaranteed to have a corresponding physical meaning, close groups of planets within the parameter space may indicate shared evolutionary pathways, and additionally shed light on the underlying workings of imputation algorithms such as the \kNNKDE which is likely finding neighbors within a cluster. Such an analysis would usually not be possible with an incomplete dataset that includes missing values.

\begin{figure*}[!ht]
    \centering
    \includegraphics[width=0.95\textwidth]{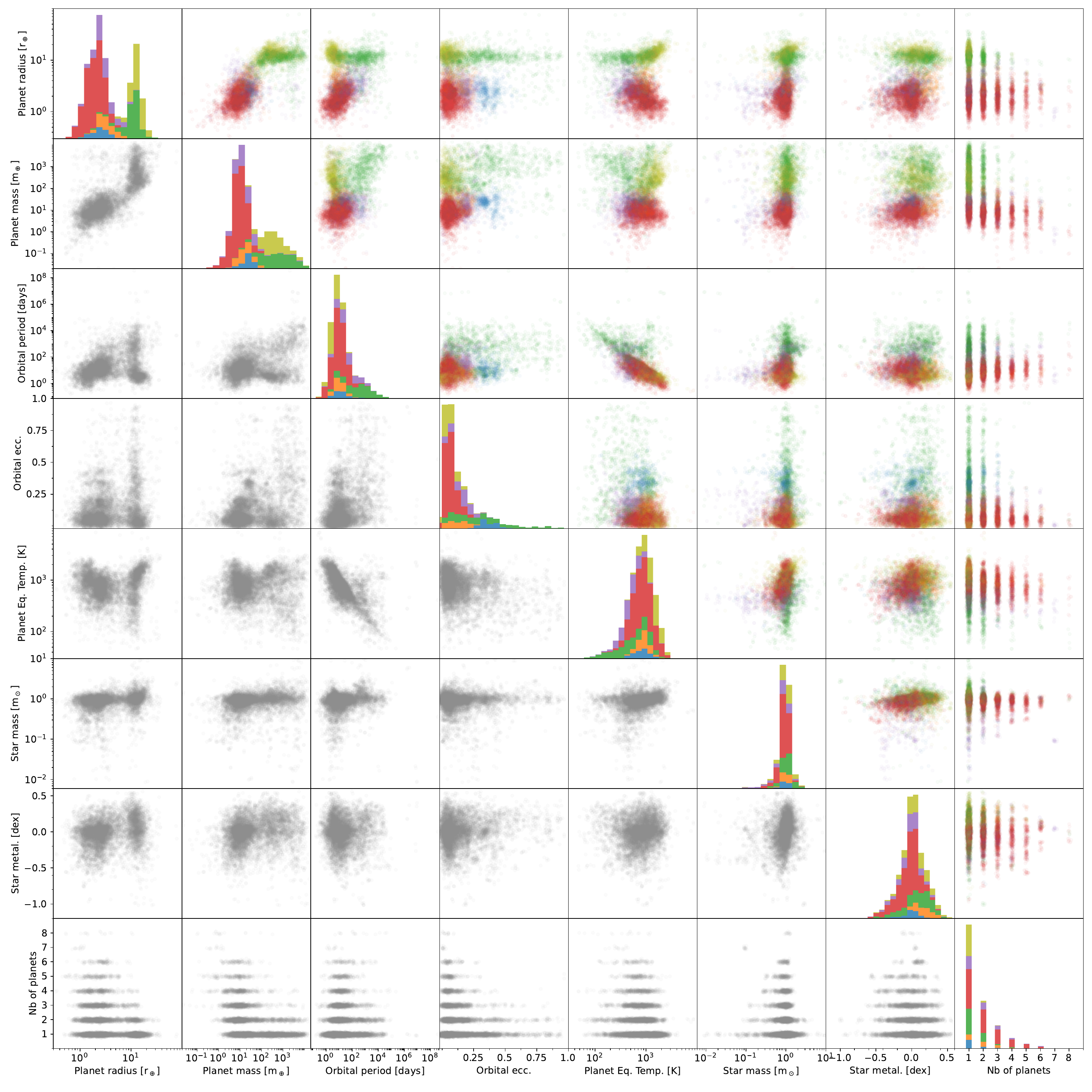}
    \caption{Pairplot for the 10,000 synthetic planets generated by the \kNNKDE. The color scheme refers to the clusters selected in Figure~\ref{fig:tsne_clusters}.}
    \label{fig:pairplot_generative}
\end{figure*}

Based on the eight property extended planet dataset, the \kNNKDE was used to create a synthetic population of 10,000 planets. The pairplot for the synthetic data looks the same in structure to Figure~\ref{fig:pairplot}, and can be seen in Figure~\ref{fig:pairplot_generative}. This large new, full properties, synthetic population was analyzed using the $t$-distributed Stochastic Neighbor Embedding ($t$-SNE) algorithm; a statistical technique that maps each datapoint in a high-dimensional parameter space to a two-dimensional space which can then easily be visualized \citep{VanDerMaaten2008}. For this analysis, the eight properties extended dataset was used, but the number of known planets in the system was discarded before passing the data to the $t$-SNE. This was because this last variable only takes discreet values and it was found to strongly dictate the resulting clusters. 

\begin{figure*}[!ht]
    \centering
    \includegraphics[width=0.5\textwidth]{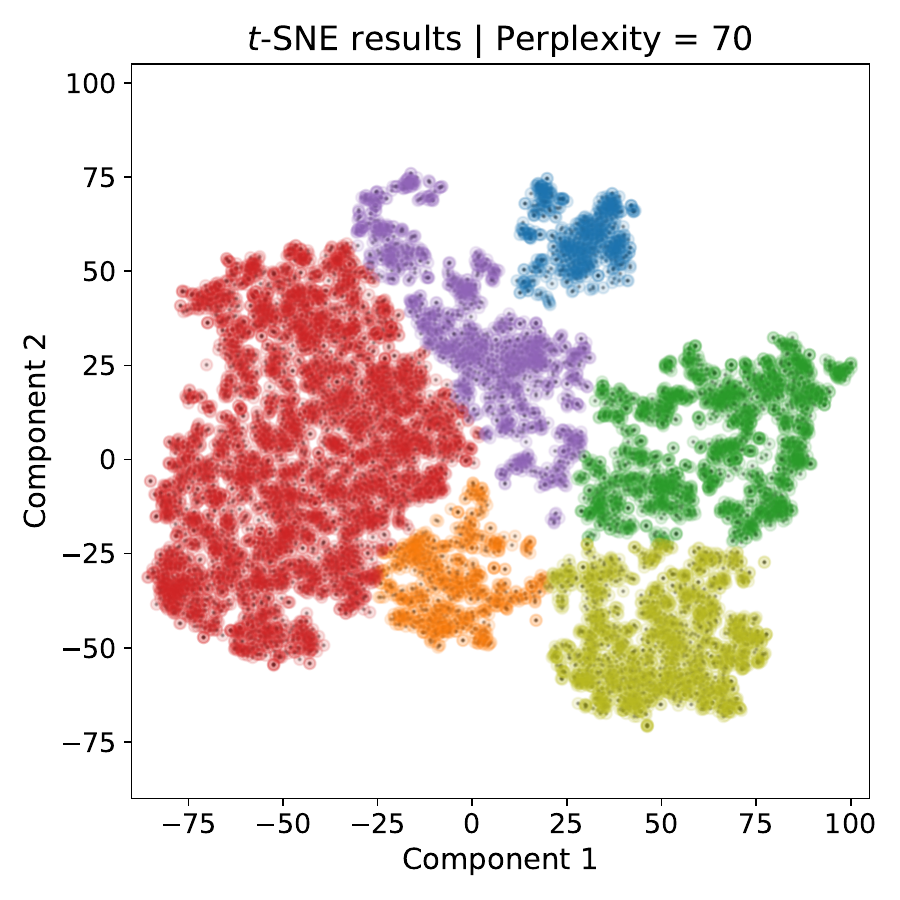}
    \caption{A 2-d visualization of clusters of planets within the seven dimensional parameter space, created using the $t$-SNE algorithm on a population of 10,000 simulated planets generated by the \kNNKDE. Six clusters identified by eye have been color-coded for subsequent analysis. The proportion of the planet population in each cluster is 42.7\,\% red, 6.9\,\% orange, 12.4\,\% purple, 5.5\,\% blue, 15.7\,\% yellow, and 16.8\,\% green. The main parameter for the $t$-SNE is the perplexity, which was set to 70. The two axes label the projected 2-d space, but have no physical meaning.}
    \label{fig:tsne_clusters}
\end{figure*}

In projecting the 7-d data onto the 2-d plane, the $t$-SNE uses non-linear transformation such that nearby points in the 2-d plane are also in the same vicinity within the original parameter space. The main parameter is the ``perplexity'' which controls the balance between global and local structure, so that a higher perplexity level typically results in more clumping. The result of the $t$-SNE is shown in Figure~\ref{fig:tsne_clusters}. The perplexity value is set to 70, although roughly the same structure is visible from perplexity values of 25 through to 200. Below a perplexity of 25, the 2-d projection of the data appears nearly homogeneous. Six clusters were identified by eye in the 2-d plane, and manually colored in Figure~\ref{fig:tsne_clusters}. It is important to note that the choice of these six clusters is arbitrary, and nothing prevents the selection of finer or coarser clusters. The selection made here includes one dominant group, two groups of intermediate size, and three smaller groups. The dominant red cluster accounts for 42.7\,\% of the 10,000 synthetic planets. The intermediate size yellow and green clusters account for 15.7\,\% and 16.8\,\% respectively. The orange, purple, and blue clusters lie in the boundary between the red dominant group on one side, and the yellow/green medium sized groups on the other, and account for 6.9\,\%, 12.4\,\%, and 5.5\,\% of planets respectively.

Analysis of the clusters in Figure~\ref{fig:tsne_clusters} reveals that the largest red cluster corresponds to the majority of the super Earths. Red cluster planets have radii below about 6\,\RE, short orbital periods and circular orbits. In Figure~\ref{fig:pairplot_generative}, they largely fill the lower left planet population in the planet mass versus radius plot. The two clusters diametrically opposite in Figure~\ref{fig:tsne_clusters} colored yellow and green are the gas giants. Both clusters contain large and massive planets but are distinguished by their orbital period and eccentricity. The upper green cluster planets have orbital periods longer than 50\,days and a wide range of eccentricities. The lower more compact cluster has short periods and low eccentricity. These yellow cluster members are the hot Jupiters, and the density of the cluster reflects a fairly uniform set of properties. The close proximity to their long orbit counterparts could support the main formation theory that hot Jupiters form through the same mechanism as cooler gas giants and migrate inwards. However, the dataset does not contain information such as composition, which would be one of best indicators of a different formation mechanism.

The orange and purple clusters lie at the transition between the red and yellow/green clusters in the 2-d plane. This also equates to size, with planets in both clusters having masses and radii in the regime of Neptune. The two clusters are differentiated by their orbital period and stellar metallicity. Orange cluster planets typically having an orbital period of about 10\,days and are found around stars of higher metallicity (greater than 0 dex). Purple cluster planets have orbital periods that peak closer to 30-40\,days and are found over a broader range of metallicities but skewed towards lower values. The two clusters are not as clearly separated as the red cluster is from the yellow and green clusters, and they may not represent distinct evolutionary pathways. If they do, it is possible that orange cluster planets are formed in more massive protoplanetary disks around higher metallicity stars and therefore migrate onto shorter orbits than their purple companions.

Finally, the small, compact blue cluster constitutes the most clearly defined cluster in Figure~\ref{fig:tsne_clusters}. Blue cluster planets have universally short and eccentric orbits at super Earth masses. In the generative pairplot in Figure~\ref{fig:pairplot_generative}, they represent the extension towards high eccentricity from the super Earth planet sizes on the planet radius versus orbital eccentricity plot. In Figure~\ref{fig:pairplot}, this group is present although not as clearly defined as the high eccentricity gas giants. But the generated synthetic planet population increases it prominence. The planets in the blue cluster also have quite high masses for their radii, giving them above average densities. These may be planets undergoing dynamical evolution, due to interaction with another planet in the system or previous scattering event. If so, the group properties do differ from other planets as the system is not yet settled.

The $t$-SNE provides a way of classifying groups of planets based on their properties in a multidimensional space. However, while some distinct clusters do exist, the properties these correspond to are clearly not unambiguous. This is similar to the shape of the probability distributions found by the \kNNKDE, which often indicated the presence of continuous range of possible values, rather than extremely distinct options. Planet classes are therefore likely also to be a continuous scale, without sharp distinctions.

\section{Discussion and Conclusions}
\label{sec:conclusions}

The NASA Exoplanet Archive is a large and invaluable source of data on the known properties for the extrasolar planet population. However, this data has been difficult to fully utilize. Due to the variety of techniques that are needed to probe the diversity of planets around other stars, the archive has many missing values, which are difficult to impute because of complex inter-dependencies between the planet and stellar properties. 

In this work, we compared five different algorithms that can impute missing values by leveraging incomplete multidimensional datasets: the $k$NN-Imputer, MICE, MissForest, GAIN and the newly developed \kNNKDE. The first four codes returned single point estimates of the missing property, while the \kNNKDE returned a probability distribution that was averaged for the algorithm comparison. These algorithms can estimate the unknown properties of a planet, based on the observed measurements of that planet and the other planets in the archive. While the algorithms could impute any missing property present in the dataset, we focused on the estimation of planet mass, as this is one of the most difficult properties to measure with observation. The algorithms are able to provide a mass estimate for any planet, regardless of which properties have measured values.

The algorithm performance was compared when using three different datasets drawn from the NASA Exoplanet Archive. The {\it complete properties dataset} consisting of 550 planets with measured values for the six properties; planet mass, planet radius, orbital period, equilibrium temperature, number of known planets in the system, and stellar mass. The {\it full archive dataset} expanded this to include an incomplete set of those six properties for all 5,251 planets in the archive. Finally, the {\it extended dataset} added two extra properties, the planet orbital eccentricity and the stellar metallicity. 

The main results from the algorithm and dataset comparison are as follows:

\begin{enumerate}
    \item When using the complete properties dataset, the overall performance from all five codes was comparable, and slightly better than the modified Boltzmann Machine (mBM) neural network presented in the previous work, TLG2020. This demonstrated that the use of algorithms capable of utilizing incomplete datasets does not degrade performance. 
    
    \item The mass imputation was tested for all algorithms by concealing the mass value for each planet with that measurement and comparing the imputed result with the observed value. This is similar to imputing mass for a planet observed with the transit technique. The average error when using the complete properties dataset for a test set of 150 planets was between $\epsilon = 0.88 - 0.97$ for all algorithms except the GAIN, corresponding to an average imputed mass within a factor of $2.4 - 2.6$ of the observed value. 

    \item When the data was expanded to the full archive dataset with incomplete values, the average errors of the mass imputation for the same 150 planet test set was slightly reduced for the \kNNKDE, MissForest, and MICE algorithms. The error range for these algorithms decreased to $\epsilon = 0.83-0.92$ (a factor of $2.3-2.5$ of the observed value). This was the hoped for result, as the full archive dataset contains a factor of ten more planets than the complete properties dataset, and therefore should provide more information to increase the accuracy of the imputation.
    
    The fact that the average error improvement was not greater is due to the increase in range of the full archive dataset, which extends from a factor of ten smaller in planet mass than the complete properties dataset to a factor of two larger. A wider range of planet properties allows more planets to imputed, even those with properties lying outside the range presented in the complete properties dataset. However, this does not increase the density in all areas of the parameter space, which would result in a lower error everywhere.
    
    Where the density of the parameter space does increase, such as for Kepler-9c, the error decreased substantially. This emphasises the value of using data from multiple detection techniques to minimise artificial gaps in the parameter space as much as possible. 

    \item A strong bias towards two average mass values for the gas giants and super Earth population was seen for the $k$NN-Imputer when using the full archive dataset. This resulted in a degredation in the average error compared to the complete properties dataset. Similar but less extreme biases were also seen for MICE and \kNNKDE, but no bias was observed for the MissForest.

    For the \kNNKDE algorithm, the origin of this bias came from averaging over the returned probability distribution to give a point estimate of the mass. This could be reduced by lowering the number of neighbours (a code hyperparameter) at the cost of a less informative distributions. A better solution is not to use the average value as the mass estimate, but to select the imputed value from a visual inspection of the probability distribution (see below). 
    
    Because this is the result of averaging the distribution, we do not consider the bias an issue for the \kNNKDE algorithm. Unfortunately, the other algorithms are point estimates and the bias cannot be avoided. 
    
    \item Based on the results above, our two most recommended algorithms for the imputation of missing values would therefore be the \kNNKDE and MissForest. The MissForest scheme uses a Random Forest for regression which has previously been lauded for tabular data, and shows no bias towards dominant planets populations. However, the ability of the \kNNKDE to return a probability distribution for the imputed value is considered the most useful.
    
    The GAIN algorithm consistently gave the worst performance due to a ``mode-collapse'' problem where the large number of high mass gas giants in the dataset caused an overestimation of the mass throughout the planet population. We do not recommend GAIN for this kind of tabular imputation.  

    \item The mass probability distributions returned by the \kNNKDE can be combined with a measurement for the planet minimum mass, yielded by radial velocity observations. In the radial velocity test for the mass imputation accuracy (where a minimum mass value is known but planet radius is an unknown), the average error was $\epsilon = 0.291$, corresponding to a mass within 1.33 of the observed value. For the same test set of planets, this reduced slightly to $\epsilon = 0.241$ (1.27 of the observed value) for the full archive dataset. The lower error indicates the importance of a minimum mass over the planet radius when estimating the planet mass. 

    \item The extended dataset was used only for the \kNNKDE algorithm. The addition of the extra two properties again slightly decreased the error for the mass imputation for the 150 planet test set, from $\epsilon = 0.86$ (full archive dataset) to $0.84$ (a factor of 2.3 of the observed value). Examination of the individual planets revealed that the inclusion of the extra properties did significantly reduce the error in the case where those properties had high values, as the value helped identify specific planets whose properties guided the mass estimate. However, the \kNNKDE algorithm weights all properties equally when searching for neighbors, and so uninformative properties can dilute more meaningful values. The potential value of the extra data should therefore be considered when imputing with the \kNNKDE, using guides to its importance such as the pairplot in Figure~\ref{fig:pairplot}.
   
\end{enumerate}

The probability distributions returned by the \kNNKDE algorithm allow the possibility to work backwards from the mass imputation to the trends being leveraged in the underlying dataset. Understanding the origin of the result (often a challenge with machine learning on complex datasets) enables to assess its reliability, and also reveals information about the demographics of the neighboring planet population in the parameter space. In this paper, we analysed 25 planet mass distributions for high and low error planets. While a thorough analysis in comparison with a data plot such as the pairplot in Figure~\ref{fig:pairplot} will offer the most information about a mass imputation, there are some ``rules of thumb'' that can give a quick insight into a distribution:

\begin{enumerate}
    \item A narrow profile such as HAT-P-57b suggests that the algorithm is confident about the mass prediction. This usually means that the neighbors in the multidimensional parameter space to the planet's observed values have a narrow range of masses and belong to a consistent group, such as the hot Jupiters. It can also mean that the planet is a particularly close match to just a small number of known planets, as in the case of TRAPPIST-1f.
    
    \item A broad or flat profile such as that of TIC 172900988b is the clearest indication that the algorithm is not confident about the mass prediction. The neighbors in the parameter space to the planet's observed values have a wide range of masses and do not seem to belong to any consistent group. This most commonly occurs when a planet lies in a sparsely populated region of the parameter space. 

    \item Multiple peaks in the distribution indicate that different mass categories of planet (such as rocky, super Earths or gas giants) share similar properties to the observed values of the planet whose mass is being imputed. 

    This can arise when a wide range of masses are compatible at that point in the parameter space (e.g. Kepler-30c), or because the planet sits in a sparse parameter area close to denser regions, such as the inflated Kepler-9c, which was sandwiched between the super Earths and gas giants.  

    Note that in these cases, it does not make sense to use the average mass as the imputed value. The peaks represent different choices, with the highest peak indicating the algorithm preferred option. One extreme case where this should be overridden is USco CTIO 108b. The very small higher mass peak is the right value. It is poorly favored by the code as there are only a few planets consistent with that mass, but it is guess-ably correct as the planet has been discovered by direct imaging which strongly favors more massive worlds. 

    In the case where a minimum mass is known, convolution with the minimum mass as described in section~\ref{sec:method_minmass} can break a multi-peak degeneracy and select the best option.

    \item A known minimum mass value that falls outside the main body of the mass distribution predicted by the code is also indicative of an error, such as in K2-111b in Figure~\ref{fig:cd_rv_profiles} (prior to the discovery of an extra planet in the system). Although the convolution step described in section~\ref{sec:method_minmass} can combine the two results, the final mass is unlikely to be accurate if the overlap between the two distributions is minimal. 
  
\end{enumerate}

In addition to assessing the accuracy of the mass imputation, study of the shape of different distributions can reveal information about the planet discovery, and the demographics of the observed planet population. One interesting example was that the very high error for the mass of K2-111b when using the complete properties dataset was due to a undiscovered planet in the same system. Once that planet was included in the data given to the code, the mass estimate dramatically improved, dropping from $\epsilon = 1.57$ to $0.02$. Creating probability distributions for planets of interest, and identifying the ones that are high error, could therefore indicate planets whose observations are not yet telling the complete story. This might be useful for follow-up observation programs. 

The expansion of the dataset from the complete properties dataset through to the extended dataset also often results in broader multi-peak profiles, rather than more distinct peaks (for example, Kepler-9c and Kepler-30c). This is due to more planets being discovered at intermediate sizes, and it points to a more continuous range of properties for planets, rather than distinct classes. 

In the final section~\ref{sec:generative} of this paper, the \kNNKDE was also used as a generative model, and a cluster analysis was performed on the synthetic population of planets. Of the six clusters identified, several groups were consistent with established planet classes, including the hot Jupiters and the super Earths. The remaining four identified groups were cooler gas giants on elliptical orbits, super Earths on short but elliptical orbits, and two classes of Neptune-sized planets, one around high metallicity stars with short orbit periods, and those on slightly longer orbits. Such clusters could possibly indicate distinct evolutionary pathways for planets.

\paragraph{A final note on the \kNNKDE}

This paper was designed to explore the algorithms that could be used to leverage the complete exoplanet archive as ultimately, we want to be able to utilise all the data we have collected. After a comparison of five different algorithms, this paper favored the \kNNKDE for imputing planetary mass. The \kNNKDE is a statistics-driven algorithm that equally weights the properties in the dataset. A strength of this technique is that the origin of the distribution structure can be fairly easily understood in conjunction with 2D plots such as the pairplot in Figure~\ref{fig:pairplot}. This is an advantage over more opaque schemes such as the mBM previously presented in TLG2020. Although the mBM was limited by the necessity to train on complete properties datasets, the algorithm could also return a probability distribution. But the origin of the mBM distributions were harder to understand due to the internal feature detection and weighting employed by artificial neural networks. On the other hand, the lack of weighting of properties does have limitations. For example, if properties are added to the dataset that have a weak or non-existent relationship to other planet properties, then the proximity of these values within the parameter space will be the metaphorical ``red herring" and could cause the \kNNKDE to select less informative neighbors for the missing property distribution. This prevents feeding every parameter available in the NASA Exoplanet Archive to the \kNNKDE. Similarly, properties with expected overlap of information (such as orbital period and equilibrium temperature) might be given disproportionate importance. Future developments for imputation methods could try to address this issue by considering adaptive metrics for the nearest-neighbor search algorithm. Learning an optimal metric would enable weighting the planet properties according to their relevance in the imputation of other properties, while still allowing for interpretation of the resulting distribution. 

The information in the \kNNKDE probability distributions can be maximised by removing the neighbor cap that stops the algorithm from considering the properties of more than 20 neighboring planets in the parameter space. As the \kNNKDE weights neighbors by proximity, more distant neighbors can be informative about possible but less probable values, which can be useful such in the case of USco CTIO 108b. As mentioned above, averaging over the resultant distribution can result in a average bias. However, if a single point imputation is not required, or manually selected, this is not an issue.

\begin{acknowledgments}
This research has made use of the NASA Exoplanet Archive, which is operated by the California Institute of Technology, under contract with the National Aeronautics and Space Administration under the Exoplanet Exploration Program. The authors thank Nicholas Guttenberg and Matthieu Laneuville for their helpful comments for this research. 
\end{acknowledgments}

\bibliography{lalande_tasker_doya_2024}{}
\bibliographystyle{aasjournal}

\end{document}